# Lunar University Network for Astrophysics Research:
## Comprehensive Report to
## The NASA Lunar Science Institute
### March 1, 2012

*Principal Investigator:* Jack Burns, University of Colorado Boulder
*Deputy Principal Investigator:* Joseph Lazio, JPL

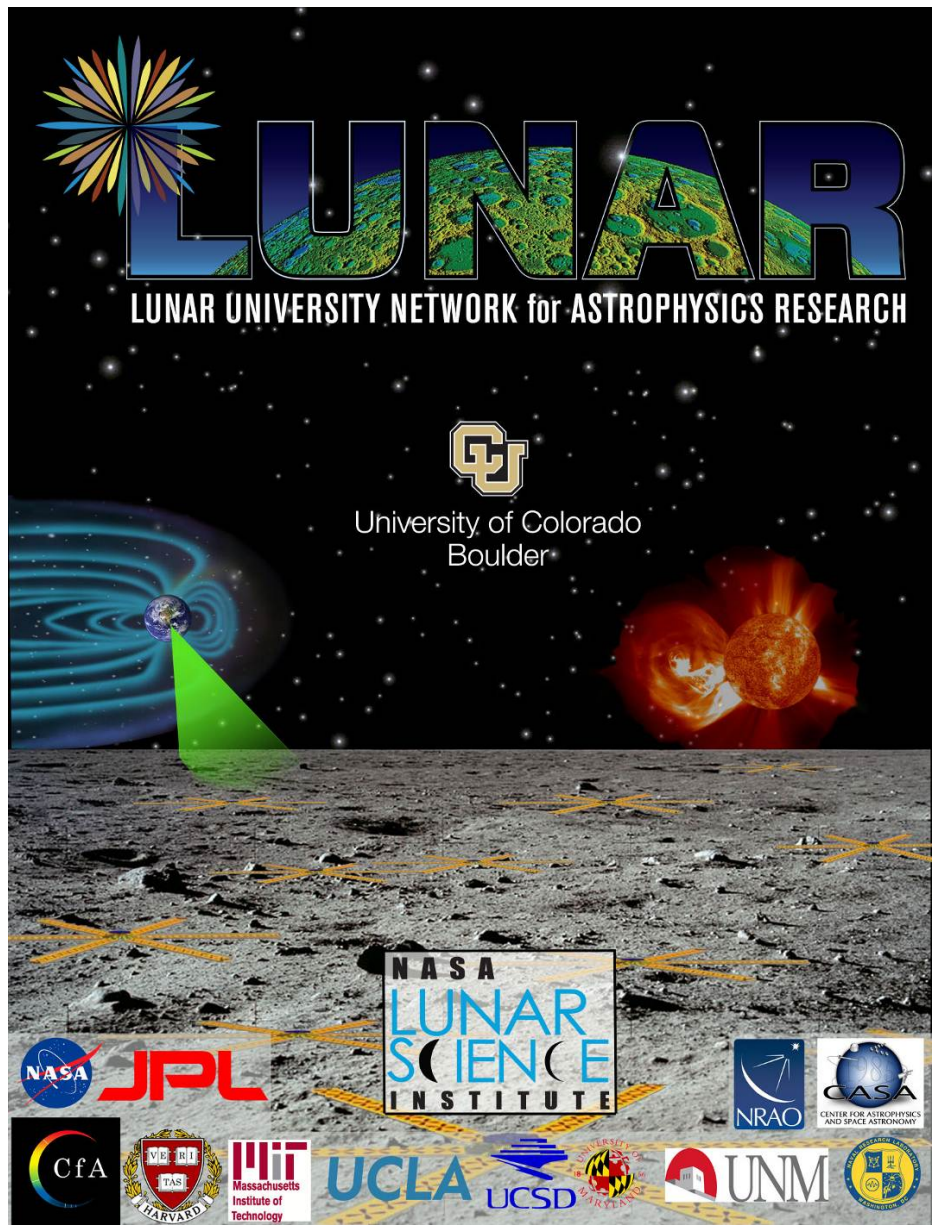

# 3.1 EXECUTIVE SUMMARY

The Lunar University Network for Astrophysics Research (LUNAR) is a team of researchers and students at leading universities, NASA centers, and federal research laboratories undertaking investigations aimed at using the Moon as a platform for space science. LUNAR research includes Lunar Interior Physics & Gravitation using Lunar Laser Ranging (LLR), Low Frequency Cosmology and Astrophysics (LFCA), Planetary Science and the Lunar Ionosphere, Radio Heliophysics, and Exploration Science. The LUNAR team is exploring technologies that are likely to have a dual purpose, serving both exploration and science. There is a certain degree of commonality in much of LUNAR's research. Specifically, the technology development for a lunar radio telescope involves elements from LFCA, Heliophysics, Exploration Science, and Planetary Science; similarly the drilling technology developed for LLR applies broadly to both Exploration and Lunar Science.

## Lunar Laser Ranging

LUNAR has developed a concept for the next generation of Lunar Laser Ranging (LLR) retroreflector. To date, the use of the Apollo arrays continues to provide state-of-the-art science, showing a lifetime of >40 yrs. This program has determined properties of the lunar interior, discovered the liquid core, which has now been confirmed by seismometry, and most of the best tests of General Relativity (GR).

> "A new Lunar Laser Ranging (LLR) program, if conducted as a low cost robotic mission or an add-on to a manned mission to the Moon, offers a promising and cost-effective way to test general relativity and other theories of gravity…The installation of new LLR retroreflectors to replace the 40 year old ones might provide such an opportunity". *New Worlds, New Horizons in Astronomy & Astrophysics (NWNH)*

However, the single shot ranging accuracy is now limited by the structure of the Apollo arrays. The next generation LLR program will provide lunar emplacements that will support an improvement in the ranging accuracy, and thus the lunar physics, by factors of 10-100.

> "Deploying a global, long-lived network of geophysical instruments on the surface of the Moon to understand the nature and evolution of the lunar interior from the crust to the core…to determining the initial composition of the Moon and the Earth-Moon system, understanding early differentiation processes that occurred in the planets of the inner solar system". *Vision and Voyages for Planetary Science in the Decade 2013-2022*

In the near term, the LLR stations that have ranged to the Apollo retroreflectors will see an improvement in accuracy of 3-11 using their existing hardware. More important, the number of returns required to obtain a 1-mm normal point is reduced by a factor of 10-100. This means that as soon as the next generation retroreflectors are deployed, the improvements in the lunar and gravitational physics will begin.

The LUNAR team has shown that the accumulation of dust on the lunar retroreflectors causes a significant loss in the return signal. Because of this, dust-mitigation techniques for use with corner cubes were studied. One such technique is to apply a hydrophobic surface coating. This coating, known as LOTUS, was originally developed to keep surfaces dust-free for missions to the Moon and Mars. The LOTUS coating is being applied to some of the corner cubes and its far-field pattern is being studied to determine the effects of the coating on the corner cube.

## Low Frequency Cosmology and Astrophysics (LFCA)

The focus of the LUNAR LFCA research is to strengthen the science case and develop relevant technologies related to tracking the transition of the intergalactic medium (IGM) from a neutral to ionized state during the time that the first stars and first accreting black holes were forming using the redshifted 21-cm signal from neutral hydrogen. The eventual goal is to exploit the "radio-quiet" properties of the Moon's farside as the site for a lunar radio telescope to conduct these fundamental measurements.

> "What were the first objects to light up the Universe and when did they do it?" *NWNH*

The Astronomy and Astrophysics Decadal Survey (*NWNH*) identified "Cosmic Dawn" as one of the three objectives guiding the science program for this decade. In the science program articulated in *NWNH* (Chapter 2), *Cosmic Dawn* was identified as a science frontier discovery area that could provide the opportunity for "transformational comprehension, i.e., discovery." This is one of LUNAR's principal scientific thrusts.

LUNAR members have been pioneers in the development of the theory and numerical simulation of the first stars and galaxies, and predictions of sky-averaged spectrum and the spatial structure of the 21-cm signal. Concurrently, LUNAR has developed new technologies and mission concepts to observe the low frequency radio signals from Cosmic Dawn. One exciting advanced concept called DARE (Dark Ages Radio Explorer) will orbit the Moon, take data only above the lunar farside, and will make the first

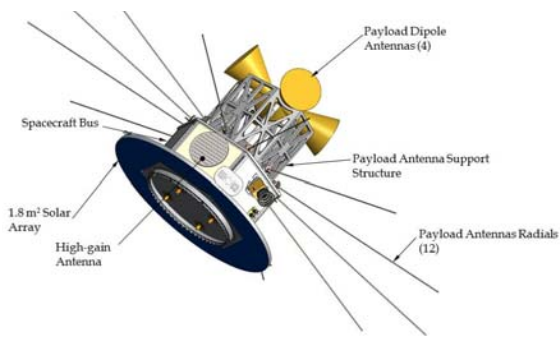

Artist's concept of DARE spacecraft.

observations of the first stars and galaxies at times <0.5 billion years after the Big Bang. In addition, LUNAR is engaged in technology development specifically to prove new antenna designs capable of being deployed in significant numbers on the lunar surface as an interferometric array. Innovative concepts include roll-out arrays of polyimide (e.g., Kapton) film with embedded metallic dipoles and magnetic helical antennas made of memory metals.

Although the primary focus of a future lunar radio telescope is likely to be Cosmic Dawn, such a telescope would be a powerful instrument for other high priority studies or would be able to conduct other interesting studies by virtue of the Cosmic Dawn observations. Examples include both searches for the magnetospheric emission from extrasolar planets and surveys for radio transients. These respond to recommendations from *NWNH*, which identified both "identification and characterization of nearby habitable exoplanets" and "time domain astronomy" as other science frontier discovery areas.

## Planetary Science connection to LFCA

*The Scientific Context for the Exploration of the Moon* (SCEM) identifies the "Lunar Environment," particularly the fact that the lunar atmosphere presents the nearest example of a surface boundary exosphere, as one of four guiding themes for science-based exploration. From this theme, the report develops a set of science goals, including "Determine the global density, composition, and

> "Planetary exospheres,…tenuous atmospheres that exist on many planetary bodies, including the Moon, Mercury, asteroids, and some of the satellites of the giant planets, are poorly understood at present. Insight into how they form, evolve, and interact with the space environment would greatly benefit from comparisons of such structures on a diversity of bodies." *Vision & Voyages for Planetary Science in the Decade 2013–2022* time variability of the fragile lunar atmosphere before it is perturbed by further human activity." The SCEM report also notes that the Moon may continue to outgas and that the lunar atmosphere, as it is coupled to the solar wind, is a dynamic system. As such, long-term monitoring is required to understand its properties. Further, as a surface boundary exosphere, studies of the Moon are likely to inform processes occurring on Mercury, other moons, asteroids, and potentially even Kuiper Belt objects. LUNAR has developed a concept to measure the Moon's ionospheric density using the plasma frequency cutoff from observations with low frequency dipole antennas on the lunar surface.

## Radio Heliophysics

High-energy particle acceleration occurs in diverse astrophysical environments including the Sun and other stars, supernovae, black holes, and quasars. A fundamental problem is understanding the mechanisms and sites of this acceleration, in particular the roles of

> "The Moon offers a large, stable surface in which to build a large, capable low-frequency radio array for the purpose of imaging solar sources at wavelengths that cannot be observed from the ground, an array that is well beyond the current state of the art for antennas in space." *SCEM* shock waves and magnetic reconnection. Within the inner heliosphere, solar flares and shocks driven by coronal mass ejections (CMEs) are efficient particle accelerators which can be readily studied by remote observations. There remain significant questions to answer about these radio bursts and the acceleration processes that produce them, including where within the CME Type II emission is produced, and how the alignment between the shock surface and the coronal magnetic field changes the acceleration. Electron densities in the outer corona and inner heliosphere yield emission frequencies below ~10 MHz. Observations must be conducted from space because the terrestrial ionosphere is opaque in this frequency range, preventing any of this emission from reaching a receiver on Earth. Work on the *Radio Observatory on the Lunar Surface for Solar Studies (ROLSS)* concept has included refinement of instrument performance requirements, prototyping of critical components such as antennas and correlator electronics, and the use of simulations and observations from analogous instruments in space or radio arrays on Earth.

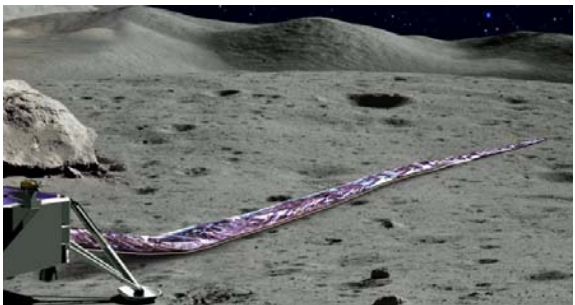

Since the first space exploration missions, in-situ interplanetary dust detection has been an important issue, both in order to understand the interplanetary and interstellar sources of dust, but also because of the effects that fast-moving dust can have on equipment and humans in space and on the lunar surface. Most dust measurements have been performed with instruments specifically designed to characterize dust particles, but recent work has

Artist's concept of a lander with a short length of polyimide film including two deposited dipole antennas (ROLSS prototype).

shown that radio receivers are also able to measure electric signals associated with individual dust grains impacting spacecraft at high speed. Based on this recent work, the LUNAR team realized that the detection and monitoring of interplanetary dust could be a valuable additional science goal for ROLSS and any future far-side low frequency radio arrays. In order to learn more about this technique, and to demonstrate that ROLSS could also make such measurements, the LUNAR team investigated dust-related signals recorded by the S/WAVES radio instrument onboard the two STEREO spacecraft near 1 A.U. during the period 2007-2010.

## Exploration Science

The LUNAR team is investigating human missions to the lunar L2/Farside point that could be a proving ground for future missions to deep space while also overseeing scientifically important investigations. On an L2 mission, the astronauts would travel 15% farther from Earth than did the Apollo astronauts and spend almost three times longer in deep space. Such missions would validate the Orion Multi-Purpose Crew Vehicle's life support systems for shorter durations, would demonstrate the high-speed reentry capability needed for return from deep space, and would measure astronauts' radiation dose from cosmic rays and solar flares to verify that Orion provides sufficient protection. On such missions, the astronauts could teleoperate landers and rovers, which would obtain samples from the geologically interesting (and unexplored) farside (i.e., South Pole-Aitken Basin) and deploy a lunar radio telescope. Such telerobotic oversight would also demonstrate capability for future, more complex deep space missions.

The LUNAR Simulation Laboratory at U. Colorado has been developed to mimic the temperature and photon radiation environment of the Moon's surface over the course of a full lunar rotation. It has tested science equipment (e.g., polyimide film antennas) and deployment techniques using mini-rovers that are envisioned for both robotic and human exploration of the Moon. It serves as a facility to test other, Exploration-specific technologies, both for the Moon or other airless bodies such as NEOs.

> "The lunar surface offers extraordinarily radio-quiet sites on the lunar farside that could enable a highly sensitive low-frequency radio telescope…An innovative concept recently proposed would have a complete antenna line electrodeposited on a long strip of polyimide film…As a result of the high astronomical priority of this work and the uniquely enabling character of the radio-quiet farside lunar surface, such efforts deserve cultivation." *SCEM*

Automated sensor deployment techniques are needed for surface missions in which instrument packages (either for Exploration or science) should or must be deployed at some distance from a lander. One option for deployment is a rover, but lower mass options, involving spring-and-pulley systems, are also being explored by the LUNAR team. Intended originally for lunar sensors, a spring-and-pulley system may be a useful technology for other low gravity environments.

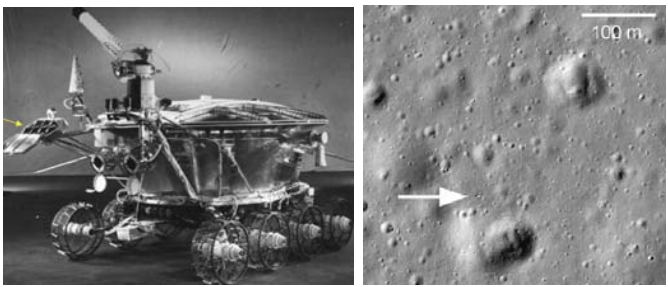

*Left:* Lunokhod 1 lander. *Right:* LRO image of Lunokhod 1.

An accurate grid or selenographic coordinate system will be critical in future robotic and/or manned missions. Lunokhod 1 had never been ranged to since the coordinates of the final resting place were not sufficiently accurate. Using the high resolution imagery, LRO identified the location of Lunokhod 1 and these coordinates were given to the LUNAR team for use at the Apache Point Observatory Station. This allowed laser returns to be obtained from Lunokhod 1 for the first time. However, the LLR position of Lunokhod 1 was different from the LRO coordinates by 100 meters. Thus, this new data point should allow a very significant upgrade to the seleodetic coordinate system being used by LRO. Additional retroreflectors will serve to tie down the coordinate system in a variety of new locations.

Drilling in the lunar regolith requires a different approach than used on Earth, as the Apollo astronauts discovered. Such drilling may be required either for stability or thermal control. The LUNAR team, with partner Honeybee Robotics, has been exploring a gas-assisted pneumatic drill, which is demonstrating relatively deep penetrations with limited resource requirements.

Regolith may prove to be an effective construction material. The LUNAR team has been exploring how modest equipment could be used to fuse lunar regolith into a concrete-like material, which could then be used for construction of large structures and astronomical telescope mirrors, without the expense of having to carry most of the material to the surface.

## Education & Public Outreach (EPO)

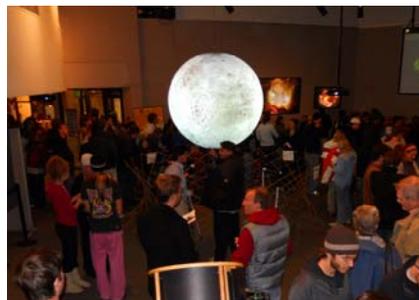

LUNAR has a diverse and aggressive EPO effort aimed at enhancing the awareness and knowledge about the Earth-Moon system. The largest elements involve the creation of a nationally-distributed children's planetarium show and extensive teacher workshops, many in partnership of the *Astronomical Society of the Pacific*. Another key element is support for high school robotics clubs making their own models of a lunar rover capable of deploying a radio telescope on the lunar surface. A final strategy is to take advantage of NASA missions and natural events such as eclipses to increase public awareness of science and of NASA's role.

The children's planetarium program is based on the award-winning book, "Max Goes to the Moon". NASA astronaut Alvin Drew played a role in the development of this show. On his mission to the ISS, he read the story "Max Goes to the Moon" to the children of Earth. Alvin introduces the story in our planetarium show. Using our well-developed process of "formative evaluation", we showed the program to test audiences of school children of the target age and also to hundreds of lunar scientists at the 2011 NLSI workshop. The feedback we gathered resulted in significant improvements to the show. "Max" is now complete and we are beginning distribution of the program.

Our numerous K-12 teacher workshops focused on getting the latest discoveries about the Earth-Moon system and cosmology and the early Universe into the classroom. By holding workshops at the Astronomical Society of the Pacific meeting, we increased the number of teachers reached to much higher numbers than we originally proposed.

We began working with high school robotics clubs to challenge them to build a small rover that could deploy low frequency radio telescopes on the nearside and farside of the Moon. A student from one of these schools was invited to the NLSI Lunar Forum in the summer of 2011 to show his rover to the lunar community.

We planned public events associated with the LCROSS/LRO mission as well as the lunar eclipse in December 2010. The lunar eclipse was the largest public astronomy event in Boulder, CO since the "Deep Impact" comet mission in 2005. Approximately 1500 people crowded into a planetarium that seats 212 (using the lobby, the grounds, and the surrounding university). All heard about NASA's lunar science in addition to seeing the eclipse.

## 3.2 LUNAR PROJECT REPORTS

### 3.2.1 LUNAR LASER RANGING: LUNAR INTERIOR PHYSICS AND GRAVITATION

**The Lunar Laser Ranging Retroreflector for the 21st Century (LLRRA-21)**
*Investigators*: D. Currie & B. Behr, U. Maryland; S. Dell'Agnello & G. Delle Monache, INFN-LNF*; K. Zacny, Honeybee Robotics

LUNAR team members led by Co-I Currie have developed a concept for the next generation of Lunar Laser Ranging (LLR) retroreflector (Currie et al. 2011). To date, the use of the Apollo arrays continues to provide state-of-the-art science, showing a lifetime of >40 years. This program has determined properties of the lunar interior, discovered the liquid core, which has now been confirmed by seismometry, and most of the best tests of General Relativity (GR). However, the single shot ranging accuracy is now limited by lunar librations and the structure of the Apollo arrays. The LLRRA-21 program will provide lunar emplacements that will support an improvement in the ranging accuracy and thus the lunar physics by factors of 10-100. The LLRRA-21 program has a very strong heritage, both in the science that has been produced using the Apollo arrays and in the retroreflector.

> "A new Lunar Laser Ranging (LLR) program, if conducted as a low cost robotic mission or an add-on to a manned mission to the Moon, offers a promising and cost-effective way to test general relativity and other theories of gravity…These are tests of the core foundational principles of general relativity. Any detected violation would require a major revision of current theoretical understanding...The installation of new LLR retroreflectors to replace the 40 year old ones might provide such an opportunity".
> *New Worlds, New Horizons in Astronomy & Astrophysics*

> "Deploying a global, long-lived network of geophysical instruments on the surface of the Moon to understand the nature and evolution of the lunar interior from the crust to the core will allow the examination of planetary differentiation that was essentially frozen in time some 3 billion to 3.5 billion years ago. Such data (e.g., … laser ranging, …) are critical to determining the initial composition of the Moon and the Earth-Moon system, understanding early differentiation processes that occurred in the planets of the inner solar system, elucidating the dynamical processes that are active during the early history of terrestrial planets, understanding the collision process that generated our unique Earth-Moon system, and exploring processes that are currently active at this stage of the Moon's heat engine".
> *Vision and Voyages for Planetary Science in the Decade 2013-2022*

In the near term, the LLR stations that have ranged to the Apollo retroreflectors will see an improvement in accuracy of 3-11, depending on the station, using their existing hardware. More important, the number of returns required to obtain a 1-mm normal point is reduced by a factor ranging from 10-100. This means that as soon as the LLRRA-21s are deployed, the improvements in the lunar and gravitational physics will begin.

We have optimized the design (see figure below), fabricated parts for thermal vacuum testing, and performed thermal/vacuum/optical tests of the current design in the SCF chamber at the INFN-LNF in Frascati, Italy, a unique facility developed for testing of retroreflector arrays

(*Dell'Agnello et al. 2010*). These efforts on testing and thermal simulation have been supported by the INFN-LNF and the ASI. With LUNAR funding, it is expected that a flight ready and certified package will be available by the end of 2012. In order to better evaluate the expected future science, the LUNAR LLR team has developed an international agreement to study the

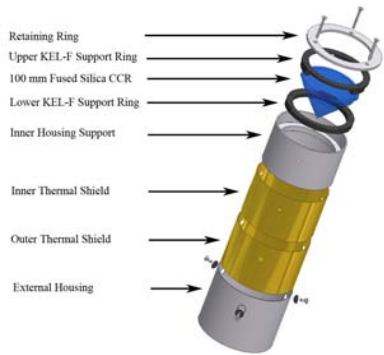

Current Design of the LLRRA-21.

future accuracy of the scientific parameters to be obtained with greater ranging accuracy. The teaming includes U. Maryland, the National Astronomical Observatory of Japan (NAOJ), JAXA, INFN-LNF, and the IAS. This has also led to a collaboration with the GRAIL team, since GRAIL will produce very accurate measurements related to the upper interior of the moon and LLR will provide very accurate information on the deep interior.

In order to provide this improvement as soon as possible, LUNAR is in discussions with several of the Google Lunar X Prize (GLXP) teams (i.e., *Moon Express*, *Astrobotoics* and *Next Giant Leap*). These teams could carry the LLRRA-21 to the moon and deploy the retroreflector within the next several years.

By a long term study of the laser ranging, one can extract properties of the lunar crust and, more important, the deep interior of the moon. For example, the use of the Apollo arrays detected the liquid core of the moon. This has been done by detailed analysis of the different frequencies in the librations, that is, the study of the "sloshing" of the liquid core. This also measured the moment of inertial of the liquid core, and its size and shape. It has also addressed the more plastic Core Mantle Boundary (CMB). This CMB extends significantly into the solid interior and provides the damping mechanism for the liquid core motion. These properties were published almost a decade ago and some of these properties have recently been confirmed by the re-analysis of Apollo seismometry. The new measurements that would be available from the LLRRA-21 would greatly improve the knowledge of this liquid core and investigate the properties of the solid inner core. The relation between the liquid core and the solid inner core will indicate the sulfur content of the deep interior which in turn will address a number of the theories of the formation of the moon.

The LLRP has revealed many other properties of the lunar interior and lunar crust. These include measurements of lunar tides, the dissipation caused by the tides and liquid core, lunar core oblateness, free librations, and more details on the search for a solid inner core. The understanding of these properties will be greatly enhanced by the deployments of the LLRRA-21 and the subsequent analyses will lead to improvement by one or two orders of magnitude depending on the method of deployment, i.e., for optimal deployment, we propose pneumatic drilling into the regolith (*Zacny et al. 2012*).

Most of the best tests of General Relativity (GR), w.r.t. other theories that attempt to explain the quantization of GR and Dark Energy, have been accomplished the LLRP. LLRRA-21 will push these evaluations one or two orders of magnitude further.

A project was initiated in order to evaluate the effect of micrometeorite impacts as an explanation of the reduced return from the Apollo arrays. This is critical in order to implement design aspects for the LLRRA-21 to assure a long lifetime with high return signal strength. To this end, witness sample plates were bombarded with accelerated dust particles in the CCLDAS dust accelerator at U. Colorado. This resulted in some cratering by dust particles of the highest energy, but given the energy vs. number information, this does not seem to be the dominant cause. This now appears to be dust, either secondary ejecta or lofted dust.

Discussions with the DREAM team to address the expected trajectories of lofted dust, but there appears to be insufficient progress in the area to directly affect the LLRRA-21 design.

A productive collaboration has been developed between U. Maryland and the INFN-LNF in Frascati, Italy. Initially, this consisted of a thermal vacuum facility explicitly designed for the optical and thermal testing of CCRs and retroreflector arrays. This resulted in the development of thermal analyses and simulations of performance for candidate LLRRA-21 designs.

The NAOJ has been working with the LLR analysis program to evaluate the parameters of the science to be achieved with the retroreflector that is proposed for SELENE-2. An agreement between U. Maryland, the NAOJ, JAXA, the INFN-LNF and the IAS has been signed to extend this to an evaluation of expected science accuracy and time scales for a 1-mm ranging accuracy from multiple ground stations.

### Advancing LLR Hollow Corner Cube Designs for Next Generation Retroreflectors
*Investigators:* Stephen Merkowitz and Alix Preston, GSFC

The majority of the first year of LUNAR at GSFC was spent refining the science case for new LLR capabilities. In collaboration with Co-I Nordtvedt, we started investigating the possibility of testing Chern Simons modified GR using LLR. We also submitted a number of white papers that positively influenced *New Worlds, New Horizons in Astronomy & Astrophysics*.

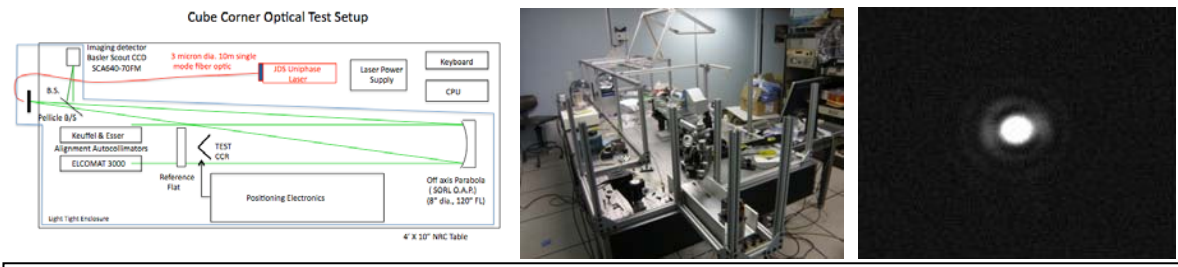

The final design of the FFDP test bed (left) along with the constructed version (center) without its enclosure, and the resultant Airy pattern from a 1arcsecond flat (right).

The LUNAR GSFC effort has recently focused on constructing a facility to design, build, and test large-scale corner cubes. Significant effort was devoted to upgrading the facilities to accommodate these needs. The first was an upgrade of the far-field diffraction pattern (FFDP) test bed. The original plan of using a transportable NRC breadboard that could be moved from lab to lab was abandoned due to instabilities in the far-field pattern and unneeded complexity of the required optics. In lieu of the transportable FFDP test bed, a more stable set-up was chosen and integrated into the laser clean room at the 1.2m Telescope Tracking Facility. The initial design provided a better far-field pattern, but vibrations and air currents appeared to be limiting the overall quality. It was also noticed that some of the existing optics could be eliminated in favor of a simpler design, as shown in figure above.

The new, simpler design was placed on a floating table to reduce seismic noise, and a plexiglass enclosure was constructed around the table to provide a more stable thermal environment. The final version of the far-field test bed can be seen in the above figure. A sample far-field pattern of a 1" diameter λ/40 flat is shown in the figure (right). Analysis of the data showed that the Airy pattern from the flat is close to the theoretical value and that each pixel covers an approximate angle of 1.5 µrad.

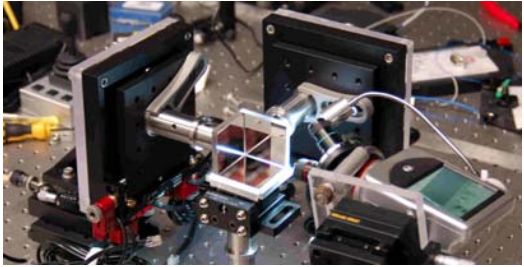

6-axis piezo aligners used to assemble hollow corner cubes. Shown are 2" mirrors being held by the aligners.

A commercial laser interferometer was purchased that will allow the angles between the facets to be easily measured at sub-arcsecond accuracies. Integration of the interferometer, software, and alignment mechanism into the Goddard facilities is currently underway. The figure to the left shows the initial assembly mechanism that was used during year-2. In this manner, the laser interferometer will be used to build the corner cubes, and the far-field test bed will be used to test the quality and performance of the corner cubes after they have been made.

In addition to upgrading the far-field test bed and integrating the laser interferometer, a significant amount of work has been done studying the best method to bond the corner cube mirrors together. To test the effectiveness of the epoxy, we plan to construct several test pieces consisting of a large window and smaller window bonded together that will be thermally cycled and then radiated. An interferometer will be used to measure the initial angle between the two flat surfaces. The samples will be thermally cycled hundreds of times and the angle between the faces will be measured again. Finally, the samples will be exposed to a radiation source to simulate decades of being unshielded on the lunar surface. The change in angle will be measured again and compared to the initial angle to determine how much the lunar environment will cause the bonds to change. This process will be done on a variety of bonding materials.

Several different bonding agents have been investigated. One method, known as hydroxide bonding, was studied in great detail. Hydroxide bonding works by placing a small amount of silicate solution between two glass surfaces. The solution then etches the glass surfaces to form siloxane chains that bond the surfaces together as the water in the solution evaporates. Tensile and sheer strength measurements between both Zerodur and fused silica have shown the bond to have strengths of several megapascals, and in some cases, pieces of glass were torn from the samples, indicating strength comparable to that of the glass itself. Although typical bond areas that have been studied are approximately 100 mm$^2$, work at Goddard has shown that this technique can be scaled up for use with the larger bond areas needed for 6" corner cubes while still retaining its strength. In addition to studying the strength of the hydroxide bond, investigations into determining how much the angles will change between the mirrors as they are bonded has also been studied. The current best estimate is that the change in angle can be kept to approximately 10-15 arcseconds from its original position. This has led us to believe that we are not being limited by the bonding technique or bonding agent itself, but by the stresses induced in the mounts as the bond is curing. We are currently working to determine how to best minimize these stresses.

## Studies of Dust and Mirror Coatings on Apollo LLR Retroreflectors

*Investigator:* Thomas Murphy, U. California at San Diego

We have shown that the accumulation of dust on the lunar retroreflectors causes a significant loss in the return signal. Because of this, dust-mitigation techniques for use with corner cubes were studied. One such technique is to apply a hydrophobic surface coating that was developed at GSFC. This coating, known as LOTUS, was originally developed to keep surfaces dust-free for missions to the Moon and Mars. A new development in coating technology has allowed us to apply a veneer at <100 C, so as to not destroy the mirror coatings. The LOTUS coating is being placed on some of the corner cubes; the resulting far-field pattern will be studied to determine the effects of the coating on the corner cube.

In addition to LOTUS, the reflective coating of the corner cubes is being studied. Although a silver or aluminum coating is typically applied to the glass facets used to construct hollow corner cubes, the extreme temperature swings that will be encountered on the Moon may cause distortions in the shape of the glass due to the large mismatch in coefficients of thermal expansion. Thus, both a metallic and dielectric coating will be used to construct several corner cubes, and the effects of each coating on the Far-field pattern will be studied as the corner cube is thermally cycled.

Efforts are underway to demonstrate definitively that the Apollo lunar reflectors suffer thermal lensing effects. Recent LLR measurements during a total lunar eclipse show that the signal response of the reflectors is dramatically altered as the sunlight is removed and then re-introduced. The total-internal-reflection corner cubes should be effective at rejecting solar energy. The most likely scenario accounting for the observations is dust deposition on the front face of the corner cube prisms. As such, it is vitally important to characterize and ultimately understand this mechanism in order to mitigate its effects in a new design.

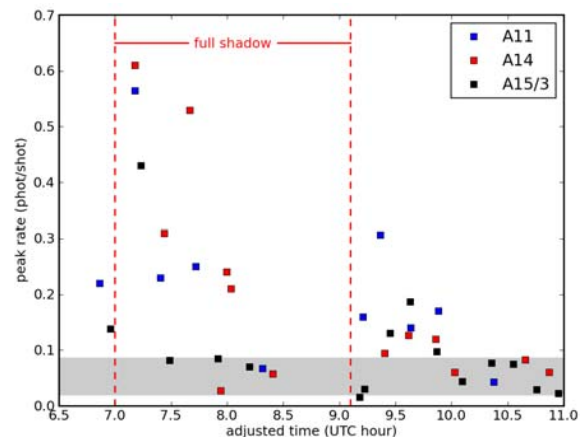

Peak photon return rate in a series of measurements to the Apollo reflectors during the total lunar eclipse of 2010 December 21. Historically, APOLLO has seen peak rates within the gray band: well short of the eclipse performance. Late in the eclipse, the signal became too weak to acquire. Performance is best during transition, as the thermal gradients in the corner cube reverse.

The eclipse performance is shown in the figure (to the right) for the three reflectors. The gray band is the historical range of performance seen at full moon for the same apparatus, so that one immediately sees something very special happening during the eclipse. In shadow, the thermal gradient reverses as the corner cube begins to radiate its thermal energy to space. When the gradient goes through zero, the response is quite good. Note a second zero-crossing when the sunlight returns and the gradient again reverses to a hot front face, where presumably dust absorbs solar energy.

## 3.2.2 LOW-FREQUENCY COSMOLOGY & ASTROPHYSICS: PROBING COSMIC DAWN

The focus of the Low Frequency Cosmology & Astrophysics Key Project is to strengthen the science case and develop relevant technologies related to tracking the transition of the intergalactic medium (IGM) from a neutral to ionized state during the time that the first stars and first accreting black holes were forming. The eventual goal is to exploit the "radio quiet" properties of the farside of the Moon as the site for a lunar radio telescope to conduct these fundamental measurements. In part, this work builds on the Lunar Radio Array (LRA) concept study performed under the Astrophysics Strategic Mission Concept Studies (ASMCS) program.

### Recommendations from *New Worlds, New Horizons in Astronomy and Astrophysics*

"What were the first objects to light up the Universe and when did they do it?" — science frontier question, Origins science theme, *New Worlds, New Horizons* Decadal Survey

The *New Worlds, New Horizons in Astronomy and Astrophysics* Decadal Survey (*NWNH*) identified "Cosmic Dawn" as one of the three science objectives guiding the science program for this decade. In the science program articulated in *NWNH* (Chapter 2), the *Epoch of Reionization* (EoR) and *Cosmic Dawn* were identified as a science frontier discovery area that could provide the opportunity for "transformational comprehension, i.e., discovery."

While our primary emphasis is on the current astronomy Decadal Survey, using the Moon as a platform for probing Cosmic Dawn via low radio frequency astronomy observations has been recognized in other NRC reports and community documents. As recent examples, both the NRC report *The Scientific Context for the Exploration of the Moon* and "The Lunar Exploration Roadmap: Exploring the Moon in the 21st Century: Themes, Goals, Objectives, Investigations, and Priorities (v. 1.1)" produced by the Lunar Exploration Analysis Group (LEAG) discuss the scientific value of a lunar radio telescope. Further, and importantly, as also discussed by *The Scientific Context for the Exploration of the Moon*, the "scientific rationale for lunar science and its goals and recommendations are independent of any particular programmatic implementation." Although the primary focus of a future lunar radio telescope is likely to be Cosmic Dawn, such a telescope would be a powerful instrument for other high priority studies or would be able to conduct other interesting studies by virtue of the Cosmic Dawn observations. Notable examples of other high priority studies would include both searches for the magnetospheric emission from extrasolar planets and surveys for radio transients. Both of these respond to recommendations from *NWNH*, which identified both "identification and characterization of nearby habitable exoplanets" and "time-domain astronomy" as other science frontier discovery areas.

### Theoretical Foundations

*Investigators:* S. Furlanetto (UCLA), A. Loeb (Harvard), J. Pritchard (Harvard, Imperial College London), J. Burns (U. Colorado), J. Bowman (ASU), D. Jones (JPL/CIT)

LUNAR team members have made fundamental contributions to understanding Cosmic Dawn, as indicated by a series of seminal papers (Furlanetto et al. 2006; Pritchard & Loeb 2008, 2010, 2011) describing various aspects of the expected highly redshifted hyperfine 21-cm line from neutral hydrogen (HI). Further, Furlanetto & Loeb are in the final stages of a book introducing researchers and students to the study of the Cosmic Dawn, entitled *The First Galaxies* and due to be published in early 2012.

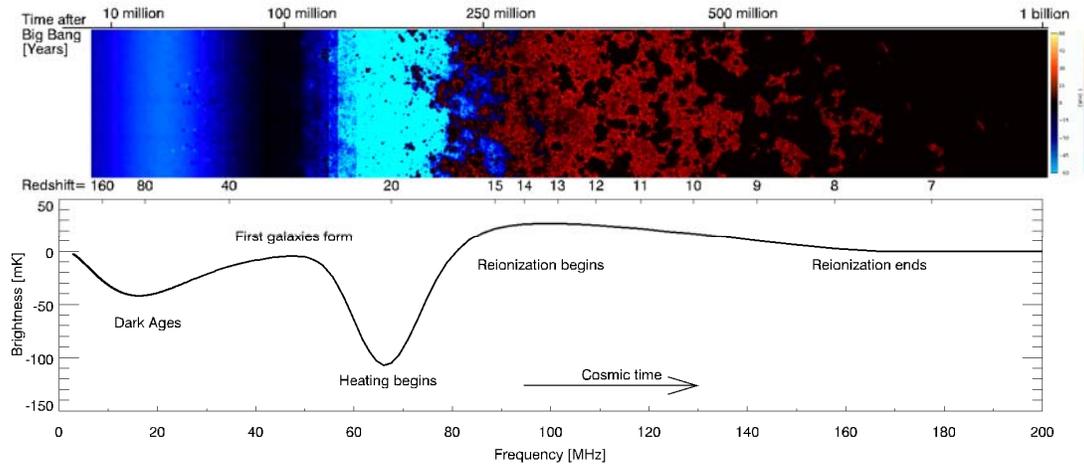

(*Top*) Time evolution of the 21 cm brightness from just before the first stars form through to the end of the EoR. Color indicates the strength of the 21 cm brightness as it transitions from absorption (blue) to emission (red) and finally disappears (black) due to ionization. (*Bottom*) Expected evolution of the sky-averaged 21 cm brightness from the "Dark Ages" at $z = 150$ to the end of the EoR sometime before $z = 6$. The frequency structure is driven by the interplay of gas heating, the coupling of gas and 21 cm temperatures, and the ionization of the gas. (Figure from Pritchard & Loeb 2010).

LUNAR team members (Bowman, Furlanetto, & Jones) also organized "The First Billion Years" workshops at the Keck Institute for Space Studies (Pasadena, CA; 2010 August and 2011 August). These workshops brought theorists, instrument builders, and observers together to consider the prospects for radio spectral signatures of the early Universe. The productive workshop triggered the large group to work together to develop new observational ideas. The principal focus was on diffuse mapping of emission lines from Cosmic Dawn, which would provide a crucial complement to probes of the 21 cm spin-flip background and help to isolate astrophysically interesting information. Lidz et al. (2011) published a paper outlining the benefits of this method, and its complementarity with low radio frequency techniques, and as a result of this workshop two instrument teams are developing plans to perform this mapping.

Specific advances and investigations occurring over the first three years of the LUNAR consortium include the following:

- Furlanetto & Johnson Stoever (2010) completed a detailed look at the fate of fast electrons in the high-redshift IGM. These are crucial to understanding the 21 cm signal that would be the focus of the LRA, and this study significantly updated past calculations with new processes and cross sections.

- Holzbauer & Furlanetto (2011) studied inhomogeneities in the UV radiation field during the era of the first galaxies. This background is crucial for "turning on" the spin-flip background that the LRA would study, and it is also crucial for modulating the transition from exotic (and very massive) Population III stars to their more normal descendants. They showed that the backgrounds are likely uniform at the 10% level, with the implication that measurements by low radio frequency telescopes will provide crucial constraints on this exotic stellar population.

- Furlanetto and LUNAR collaborator Andrei Mesinger (Princeton) developed and released the 21CMFAST semi-numeric code to predict the 21 cm spin-flip signal under a wide range of astrophysical models (Mesinger, Furlanetto, & Cen 2011). The code is now public and is being utilized by multiple groups to generate predictions for forthcoming telescopes. These tools were used by Crociani et al. (2011) to study the role of recombinations in the late phases of the EoR, one of the least understood aspects of that process (and hence of the 21 cm spin-flip signal).

- The first paper in a series by Munoz & Furlanetto (2012) compares models of faint quasars to the observations and so has implications for the heating and ionization of the IGM, a crucial input to models of the redshifted 21 cm signal.

- Furlanetto & Benjamin started to examine metal enrichment and heating of the IGM by early supernovae. In particular, they are computing the Sunyaev-Zel'dovich distortion generated by such supernovae and using it to constrain such explosions in the very early Universe. Since these explosions are a key energy input to the IGM, this exercise will improve models of the redshifted 21 cm signal.

- Loeb and collaborators completed a computer code that provides a comprehensive description of the full redshift evolution of the 21 cm signal throughout the entire cosmic history between ten million years after the Big Bang and the present time (corresponding to the redshift interval 0–100). This code allows exploration of the important scientific advantages of a lunar observatory over ground-based radio arrays.

- Loeb and collaborators explored the constraints on the EoR history that are provided by current observations of the Lyman-a forest and the microwave background. These constraints led them to the conclusion that the 21 cm signal-to-noise ratio peaks at a redshift $z$ ~ 10 and will be detectable by future experiments.

- Bittner & Loeb (2011) studied whether a modest-sized telescope, but with complete Fourier plane coverage, would be capable of constraining the EoR. They found that, even without using a full power spectrum analysis, the global redshift of the EoR, $z_{reion}$, can in principle be measured from the variance in the 21 cm signal among multiple beams as a function of frequency at a roughly 1° angular scale.

- Adshead et al. (2011) considered whether future 21 cm observations would also be able to place constraints on the inflationary era of the Universe. They found that unique constraints are unlikely to be possible. However, they also found that upcoming cosmological experiments provide an intriguing probe of physics between TeV and GUT scales by constraining the reheating history associated with any specified inflationary model, opening a window into the "primordial dark age" following the end of inflation.

- Bittner & Loeb showed that the post-recombination streaming of baryons through dark matter keeps baryons out of low mass halos ($< 10^6$ M$_\odot$) on scales of a few comoving Mpc. Using a semi-numerical code, they showed that the impact of the baryon streaming effect on the 21 cm signal during the EoR (redshifts $z$ ~ 7–20) depends strongly on the cooling scenario assumed for star formation, and the corresponding virial temperature or mass at which stars form.

- Visbal & Loeb (2011) assessed the contribution of X rays to the EoR. Popular models assume that UV photons alone are responsible for the EoR. Because X-rays have a large mean free path through the neutral IGM, they introduce partial ionization in between the sharp-edged bubbles created by UV photons. This smooth ionization component changes the power spectrum of the cosmic microwave background (CMB) temperature anisotropies.

Visbal & Loeb found that models with more than a 10% contribution from X rays produce a significantly lower power spectrum of temperature anisotropies than all the UV-only models considered.

- It is commonly thought that stars are responsible for reionizing the Universe. Wyithe et al. (2011) assessed the contribution from shocked gas associated with gravitational infall into galaxies. They found that shocks can ionize no more than a few percent of the cosmic hydrogen by $z \sim 6$. However, the small fraction of ionizing radiation produced by fast accretion shocks would be significantly more biased than that associated with stars, leading to a modification of the luminosity weighted source clustering by $\sim 10\%$. This modification of the bias may be measurable with future precision experiments using the redshifted 21 cm line to study the distribution of hydrogen during the EoR.

- Mirabel et al. (2011) considered the extent to which feedback from accreting black holes in high-mass X-ray binaries (BH-HMXBs), in addition to the UV radiation from massive stars, was an important source of heating and reionization of the IGM. They concluded that a significant fraction of the first generations of massive stars end up as BH-HMXBs. As X-ray photons are capable of producing several secondary ionizations, the ionizing power of a BH could be greater than that of its progenitor. Feedback by the large populations of BH-HMXBs heats the IGM to temperatures of $\sim 10^4$ K and maintains it in an ionized state on large distance scales. This effect has a direct impact on the properties of the faintest galaxies at high redshifts, the smallest dwarf galaxies in the local Universe, and on studies of the 21 cm signal from H I.

- Visbal & Loeb (2010) showed that it is possible to measure the clustering of galaxies by cross correlating the cumulative emission from two different spectral lines that originate at the same redshift. Through this cross correlation, one can study galaxies that are too faint to be individually resolved. This technique of intensity mapping is a promising probe of the global properties of high redshift galaxies.

- Mirocha et al. (2012) have developed a method for optimally constructing discrete spectra. They show that carefully chosen 4-bin spectra can eliminate errors associated with frequency resolution to high precision, whereas commonly used discretization schemes overestimate hydrogen ionization by factors of $\sim 2$ and underestimate temperatures by up to an order of magnitude. Applying their findings to fully 3-D radiation-hydrodynamic simulations of the early Universe, we find substantially altered H II region sizes and morphologies around first stars and black holes, and therefore a sizable impact on their associated observable signatures.

## Collateral Science

*Investigators:* J. Lazio (JPL/CIT), J. Darling (U. Colorado, Boulder), G. Taylor (UNM)

We describe as "collateral science" other investigations that the LRA would likely conduct, either by virtue of its sensitivity and frequency coverage or as part of the necessary calibration work for detecting and studying the Cosmic Dawn signal.

- Lazio, Farrell (DREAM team), and collaborators have searched for magnetospheric emissions from extrasolar planets. While no such emissions have yet been detected, magnetic fields are thought to be important in determining the habitability of terrestrial planets and such observations may be an important component of assessing any terrestrial-mass planets discovered in the solar neighborhood.

- Lazio and collaborators and Darling and collaborators have assessed what other spectral lines might be present at frequencies relevant for studying the highly redshifted 21 cm signal from H I during Cosmic Dawn. They find that there is some possibility that spectral lines from other elements or molecules, either in the Milky Way Galaxy or in foreground galaxies, may be detectable by the LRA.
- Lazio, Taylor, and collaborators have been involved in various searches for transients ("time domain astronomy") at frequencies relevant for redshifted 21 cm H I studies of Cosmic Dawn. To date, only archival transients have been found, typically on relatively long durations, but real time observations are increasingly promising, particularly with the advent of the Long Wavelength Array. Some of this work was recognized on the cover of the Astronomical Journal.

## Lunar Radio Array Concept and Technology Development

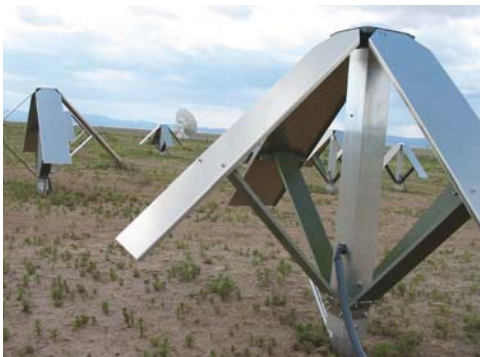

Cover photo on *Astronomical Journal* issue in which an all-sky low radio frequency survey is reported, such as might be performed by a future lunar-based telescope (Lazio et al. 2010).

The Lunar Radio Array (LRA) is a concept for a telescope located in the shielded zone on the farside of the Moon with a prime science mission of extracting cosmological and astrophysical measurements from power spectra of the highly redshifted 21 cm signal. While there are a number of on-going, ground-based efforts to detect and study the 21 cm signals from the Epoch of Reionization ($z < 15$), a combination of radio frequency interference (RFI) and ionospheric effects will make probing to higher redshifts increasingly more difficult. In many aspects, the design of the LRA will be influenced by these on-going ground-based efforts, most notably in the system architecture of the telescope and the processing of the signals. Further, there are a number of other technologies that are being developed, in many different contexts that are likely to be applicable to the LRA. Potential examples include the development of higher-energy density batteries for power storage and increased rover autonomy for telescope array deployment. One important technology unlikely to be developed in other contexts, however, is the design for the science antennas that collect the radio radiation. Moreover, the experience of the ground-based telescope arrays is of little use because those science antennas have not been developed with the mass and volume constraints of a space mission.

LUNAR team members are engaged in technology development specifically to prove new antenna designs capable of being deployed in significant numbers on the lunar surface (see figures below). Two primary technologies are being considered:

**Polyimide film dipoles**: Polyimide film is a flexible substance with a substantial heritage in space flight applications. In this concept, a conducting substance is deposited on the polyimide film to form the antenna. The film would then be rolled, for storage in a small volume during transport. Once on the Moon, the polyimide film rolls would be unrolled to deploy the antennas. Over the course of the past three years, there have been a series of increasingly complex simulations and field tests conducted on polyimide film-based antennas. Modeling and simulations have focused on two aspects of the antenna. First, is to explain the measurements, described in more detail below. Second, is to assess the electromagnetic properties of both the antenna and the anticipated transmission line for the signals. In the initial concept for the LRA, the polyimide film would also host a transmission line to carry the radio signals from the antennas to a "central electronics package." Depending upon the eventual architecture of the LRA, the central electronics package would host the receivers and data acquisition units for a few to a few hundred antennas. Analysis suggests that the transmission lines will suffer considerable losses, unless constructed from superconducting materials. An alternate approach is to have a minimum of an amplifier at the antenna feedpoints. A proof-of-concept antenna has been deployed in the field multiple times. The initial tests focused on the feedpoint impedance and understanding the frequency behavior of a polyimide film-based antenna (consisting of two 4 m long arms). Subsequent tests have included an end-to-end test in which the antenna was connected to a receiver.

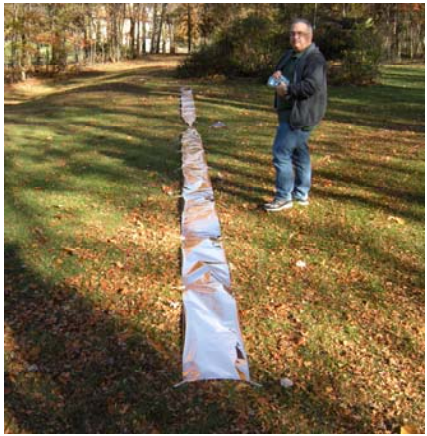 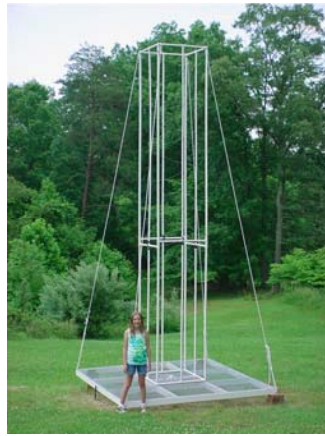

(*Left*) Polyimide film-based dipole antenna capable of being deployed by unrolling on the lunar surface. (*Right*) Magnetic helix antenna capable of being constructed from memory metals and self-deployed on the lunar surface. The antenna is wrapped around a support structure, which is required because of the high terrestrial gravity, but the antenna itself is not visible.

Simulations show good agreement between the models and the observations, suggesting that the electromagnetic properties of the antenna are being understood. Furthermore, the most recent series of tests have shown that the antenna can detect absorption due to the *terrestrial* ionosphere. While likely not important for cosmological observations, these tests indicate that a polyimide film-based antenna could also be used for lunar science. Also, while none of these tests were conducted on a lunar analog surface, the team is preparing for such tests. Finally, additional testing has also shown that these proof-of-concept antennas are likely to survive the harsh lunar environment (Stewart et al. 2011; Lazio et al. 2012).

**Magnetic helices**: In this concept, high gain antennas would be formed from helical structures. Constructed from memory metals, the antennas could be packaged in small "pizza boxes" for transport. Once on the Moon, the memory metal material would cause the antennas to assume the desired shape. Initial work focused on simulating the antenna frequency behavior and electromagnetic properties. Subsequently, a proof-of-concept antenna has been constructed, and measurements of its beam pattern have commenced. Data analysis for the beam pattern measurements is on-going.

Finally, following the concept study for the Lunar Radio Array under the Astrophysics Strategic Mission Concept Study program, one of the key issues identified was quantifying the effect of failures of antennas in an array. Taylor and collaborators conducted such studies (Taylor et al. 2011a, b). They assumed that the dipoles were distributed in a "pseudo-random" configuration that is optimized to reduce sidelobe levels. Elements were digitized to allow for independent pointings of beams and to allow for all-sky snapshot imaging. Using 919 dual polarization antennas, they found that the array is very robust against random failures of elements, but more susceptible to systematic failures.

Taylor and collaborators also took advantage of the completion of the first station of the Long Wavelength Array (LWA1) to investigate imaging the sky at 10 to 88 MHz. Their work will result in a better characterization of the emission of the Galactic background and strong radio sources. They also see in these images the strong influences of the Earth's ionosphere, which could be a major driver for a LUNAR mission.

### The Dark Ages Radio Explorer (DARE)
*Investigators:* J. Burns, Colorado; J. Lazio, JPL; S. Bale, UC Berkeley; J. Bowman, ASU; R. Bradley, NRAO; C. Carilli, NRAO; S. Furlanetto, UCLA; G. Harker, Colorado; A. Loeb, Harvard; J. Pritchard, Harvard & Imperial College London

Burns et al. (2012) have developed a concept for a new cosmology mission called the Dark Ages Radio Explorer (DARE). Considered to be a precursor for an eventual farside lunar radio telescope, DARE's science objectives include (1) When did the first stars form? (2) When did the first accreting black holes form? (3) When did Reionization begin? (4) What surprises does the end of the Dark Ages hold (e.g., dark matter decay)? DARE

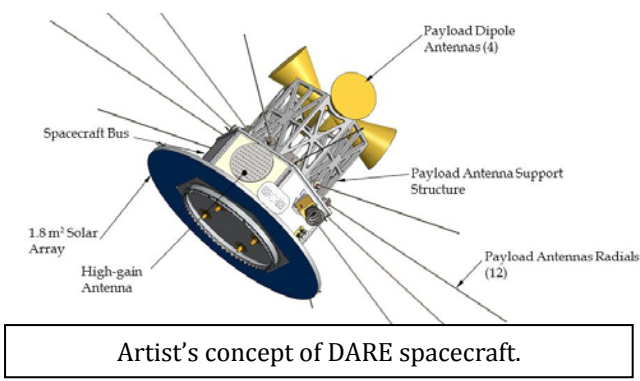

Artist's concept of DARE spacecraft.

will use the highly-redshifted hyperfine 21-cm transition from neutral hydrogen to track the formation of the first luminous objects by their impact on the IGM at redshifts 11 - 35. It will measure the sky-averaged spin temperature of neutral hydrogen at the unexplored epoch 80–420 million years after the Big Bang, providing the first evidence of the earliest stars and galaxies to illuminate the cosmos. DARE's approach is to measure the expected spectral features in the sky-averaged, redshifted 21 cm signal at 40–120 MHz. DARE orbits the Moon, taking data above the lunar farside, the only location in the inner solar system proven to be free of human-generated radio frequency interference. The science instrument is composed of a radiometer, including tapered bi-conical dipole antennas, a receiver, and a digital spectrometer.

A ground-based prototype of the DARE Science Instrument has been constructed to verify the end-to-end performance and stability including a full-scale antenna with a planar ground screen; a DARE prototype receiver using active elements (balun) with a design based on "lessons learned" from the ground-based prototype EDGES; and a digital spectrometer with 10 kHz resolution, and with radiation tolerant ADC and FPGAs available. Field tests will begin in Western Australia (at the MWA/EDGES site) in 2012 March.

We have also developed a numerical model of the DARE observations accounting for the 21-cm signal, foregrounds, and several major instrumental effects. Harker et al. (2012) applied Markov Chain Monte Carlo techniques to demonstrate the ability of these instruments to separate the 21 cm signal from foregrounds and to quantify their ability to constrain properties of the first galaxies during Cosmic Dawn.

### 3.2.3   RADIO HELIOPHYSICS

#### Particle Acceleration in the Solar Corona and Inner Heliosphere
*Investigators*: Justin Kasper, Harvard Smithsonian CFA; Bob MacDowall, NASA/GSFC

High-energy particle acceleration occurs in diverse astrophysical environments including the Sun and other stars, supernovae, black holes, and quasars.  A fundamental problem is understanding the mechanisms and sites of this acceleration, in particular the roles of shock waves and magnetic reconnection.  Within the inner heliosphere, solar flares and shocks driven by coronal mass ejections (CMEs) are efficient particle accelerators which can be readily studied by remote observations.

Solar radio bursts are one of the primary remote signatures of electron acceleration in the inner heliosphere and our focus is on two emission processes, referred to as Type II and Type III radio bursts.  Type II bursts originate from suprathermal electrons (E > 100 eV) produced at shocks.  These shocks generally are produced by CMEs as they expand into the heliosphere with Mach numbers greater than unity.  Emission from a Type II burst drops slowly in frequency as the shock moves away from the Sun into lower density regions at speeds of ~400–2000 km s$^{-1}$.  Type III bursts are generated by fast (2–20 keV) electrons from magnetic reconnection, typically due to solar flares.  As the fast electrons escape at a significant fraction of the speed of light into the heliosphere open along magnetic field lines, they produce emission that drops rapidly in frequency.  There remain significant questions to answer about these radio bursts and the acceleration processes that produce them, including where within the CME Type II emission is produced, and how the alignment between the shock surface and the coronal magnetic field changes the acceleration.  The figure below illustrates the typical structure of a CME, the possible locations of particle acceleration, and the angular resolution needed by a low frequency radio array to distinguish between acceleration locations and mechanisms.

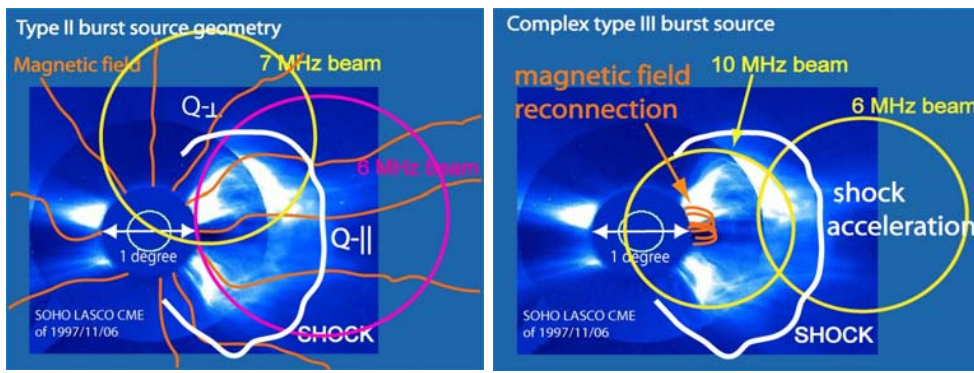

*Left:* Where on the shock does electron acceleration occur, yielding type II radio emission? *Right:* Are complex type III bursts produced by shock acceleration or reconnection?

#### A Radio Observatory on the Lunar Surface for Solar studies (ROLSS)
*Investigators*: Joseph Lazio, JPL; Bob MacDowall, NASA/GSFC; Justin Kasper, Harvard Smithsonian CfA

Electron densities in the outer corona and inner heliosphere yield emission frequencies below ~10 MHz. Observations must be conducted from space because the terrestrial ionosphere is opaque in this frequency range, preventing any of this emission from reaching a receiver on Earth. The figure to the right illustrates the active low-frequency radio environment in space, including terrestrial radio frequency interference (RFI), as seen by the WAVES instrument on the Wind spacecraft (Bougeret et al. 1995). Weak type III radio bursts are seen as nearly vertical "bars" of emission. Solar radio observations from the near side of the moon would necessarily be made in the gaps between the RFI bands. Observations of

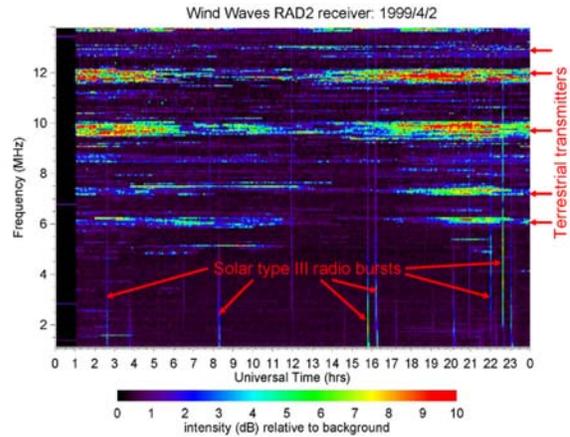

The 24-hr dynamic spectrum covers 1–14 MHz, intensity is shown by a logarithmic color scale.

faint sources require an observatory on the far side of the moon, shielded from the RFI.

> "The Moon offers a large, stable surface on which to build a large, capable low-frequency radio array for the purpose of imaging solar sources at wavelengths that cannot be observed from the ground, an array that is well beyond the current state of the art for antennas in space."
> *The Scientific Context for the Exploration of the Moon*

The Radio Observatory on the Lunar Surface for Solar studies (ROLSS) concept meets the requirements for observing solar radio bursts, which are typically more intense than those in the above dynamic spectrum. It consists of 3 arms of thin polyimide film (PF), each 500 m in length, radiating from a central hub, providing ~2° angular resolution at 30-m wavelength (10 MHz). The system is located near the lunar equator on the near side. Each arm includes 16 dipole antennas, consisting of metal deposited on the film and the transmission lines connecting them to receivers at the central hub. These arms could be deployed using a crewed or robotic rover. The data collected by the antennas are processed at the central hub and down-linked to Earth for final radio image synthesis. This antenna system is uniquely suited to the low mass and low volume requirements

| Component | Mass Estimate (kg) | ROM Cost Estimate (M$) | Details of ROM estimate |
|---|---|---|---|
| Science antenna array, including development | 90 | 10 | Primarily development cost |
| Central Electronics Package | 140 | 12 | Development cost |
| Lithium Ion Batteries | 120 | 5 | Hardware cost |
| CEP Thermal system | 25 | 3 | Development & hardware cost |
| RF/Comm System | 25 | 5 | Hardware cost |
| Solar Panel Assembly | 25 | 2 | Hardware cost |
| Deployment rover system | 50 | 20 | Hardware & software development + testing |
| Operations | n/a | 12 | Labor cost + facilities |
| Systems Engineering | n/a | 4 | Labor cost |
| Project Management | n/a | 16 | Based on costs for other missions |
| Total cost excluding LV and lander | 475 | 89 | |

for delivery to the lunar surface.

Work on the ROLSS concept has included refinement of instrument performance requirements, prototyping of critical components such as antennas and correlator electronics, and the use of simulations and observations from analogous instruments in space or radio arrays on Earth operating at higher frequencies but with similar angular resolution.

In our design trade studies for ROLSS, we assumed simple scaling relationships for power consumption by the signal correlator as a function of array capabilities such as number of antennas, frequency channels, and time resolution. There are actually many properties internal to a given Field Programmable Gate Array (FPGA) at the heart of a correlator that can be adjusted to optimize the performance of the device. After evaluating the design of the array and currently available knowledge on component performance, we determined that the most useful thing to investigate for power consumption for a lunar radio array is the actual performance of FPGA-based astronomical correlators. Large N ultra low power correlators are required for lunar applications, such as the Dark Ages Lunar Interferometer (DALI), along with ROLSS. In this study, we used current terrestrial technology FPGAs (Virtex 5 family), noting that Xilinx has a contract to radiation harden the Virtex 5. This means that the Virtex 5 is a very likely candidate for any lunar radio array within the next decade. Our objectives were to examine the relationship between total power consumption and the correlator architecture, number of stations, bandwidth, correlator bit-width, and clock rate. The Figure below shows the test setup, and an example of measurements that confirmed that available FPGAs could easily provide the performance needed for ROLSS.

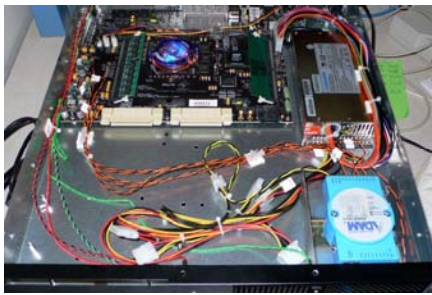 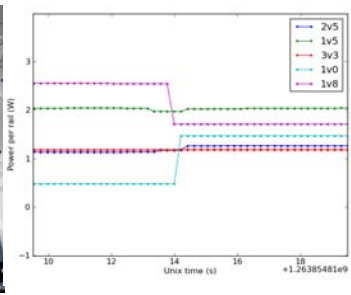

Left: A Xilinx Virtex 5 Reconfigurable Open Architecture Computing Hardware (ROACH) FPGA board was used to demonstrate that a signal correlator for ROLSS could be built using, commercially available, flight-qualified parts. Right: Measurements of the power consumption of a 15-element baseline FX correlator for ROLSS as it was enabled and then disabled.

A white paper describing ROLSS was submitted to the Solar and Space Physics Decadal Review (MacDowall et al, 2010). In that paper, we provided rough estimates of hardware costs (and mass) for such an array, excluding the costs of launch vehicle and lander/deployer (see Table above). ROLSS has been presented to several conferences, and the bulk of the work to date has been published (Lazio et al., 2011).

## ROLSS-Pathfinder

*Investigators*: Bob MacDowall, NASA/GSFC; Joseph Lazio, JPL; Justin Kasper, CfA

Although ROLSS is a relatively simple observatory to deploy and operate on Earth, the lunar environment presents many challenges. It is highly desirable to test the ROLSS concept with a pathfinder deployment of a short section of the antenna film as shown in the figure (right). We anticipate that a Google X-prize or other lander would be able to accomplish such a test

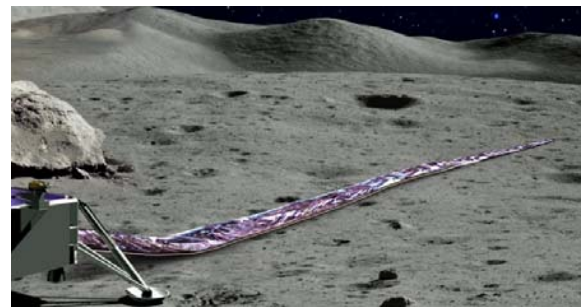

Artist concept of a lander with a short length of polyimide film including two deposited dipole antennas. (Courtesy JPL)

deployment with limited data acquisition and transmission to Earth. Such data could measure the maximum lunar ionospheric density.

We are currently developing version 2 of such a deployer (see also Section 3.2.4). This version will permit demonstration of automated film deployment and eventually of acquisition of radio data. We anticipate that a single length of film of less than 50 m with two antennas would be the right size for this pathfinder. Such a short length of film could be launched from a box mounted on the side of a lunar lander, as suggested by the figure below (left). This deployer, using a spring-launched anchor to hold the pulley for a line that would pull out the film is currently being designed and developed at GSFC. Numerous deployments of the film will be made to validate the design, after which a more flight-like model will be built. This activity was presented by MacDowall et al. (2011) at the 2011 Fall AGU meeting.

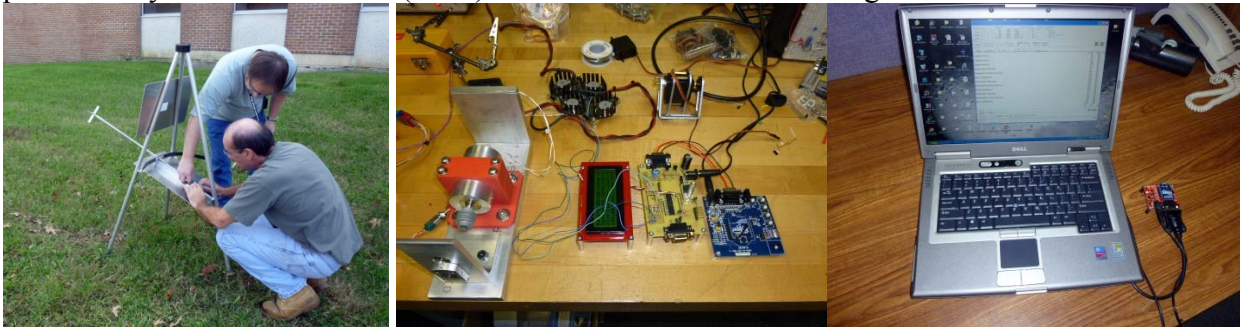

Left: Technicians ready the anchor for a test launch. Center: Electronics for remote deployment of the film. Right: Remote telemetry station.

### Dust Observations with Lunar Polyimide Film Antennas

*Investigator*: Justin Kasper & Arnaud Zaslavsky, Harvard-Smithsonian Center for Astrophysics

Since the first space exploration missions, in-situ interplanetary dust detection has been an important issue, both in order to understand the interplanetary and interstellar sources of dust, but also because of the effects that fast-moving dust can have on equipment and humans in space and on the lunar surface. Most dust measurements have been performed with instruments specifically designed to characterize dust particles (Grun et al., 1992; Srama et al., 2004), but recent work has shown that radio receivers are also able to measure electric signals associated with individual dust grains impacting spacecraft at high speed (see e.g. (Meyer-Vernet, 2001) and references therein).

Based on this recent work, the LUNAR team realized that the detection and monitoring of interplanetary dust could be a valuable additional science goal for ROLSS and any future farside low frequency radio arrays. In order to learn more about this technique, and to demonstrate that ROLSS could also make such measurements, the LUNAR team investigated dust-related signals recorded by the S/WAVES radio instrument onboard the two STEREO spacecraft near 1 A.U. during the period 2007-2010. The impact of a dust particle on a spacecraft produces a plasma cloud whose associated electric field can be detected by on-board electric antennas (shown in Figure below). For this study we use the electric potential time series recorded by the waveform sampler of the instrument. The high time resolution and long sampling times of this measurement enable us to deduce considerably more information than in previous studies based on the dynamic power spectra provided by the same instrument or by radio instruments onboard other spacecraft. The large detection area compared to conventional dust detectors provides flux data with a better statistics.

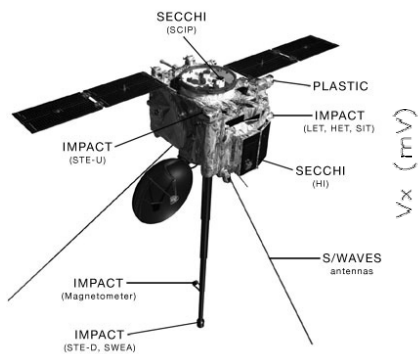

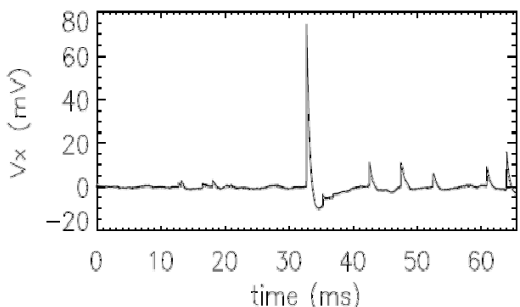

*Left:* The twin STEREO spacecraft. We have been examining the radio signatures of dust impacts. *Right:* Sudden changes to the spacecraft potential as measured by individual antennas are produced when an interplanetary dust particle impacts the spacecraft.

We showed that the dust-generated signals are of two kinds, corresponding to impacts of dust from distinctly different mass ranges, and derived calibration formulas for these signals in the nanometer and micrometer size ranges. For micrometer-scale dust, the orbital motion of the spacecraft enables us to distinguish between interstellar and interplanetary dust components (Figure to the right). These results have been presented at conferences and submitted for publication (Zaslavsky et al., 2011).

Future work on dust detection with ROLSS is focusing on the frequency dependence of these signals, and on changes to the design of the ROLSS electronics, including frequency response and electrical grounding, needed for optimum dust detection and characterization.

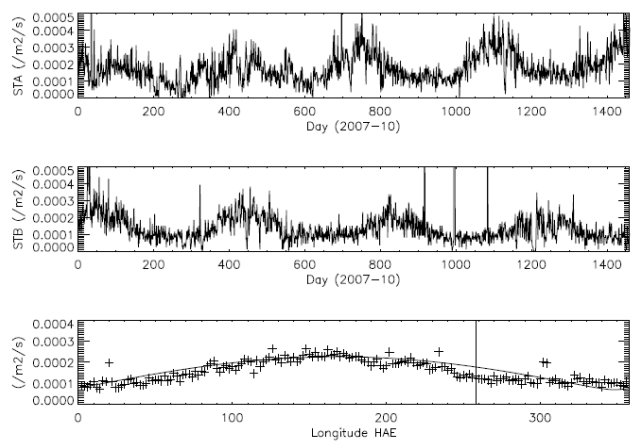

Micron-scale dust flux calculated each day for STEREO A and B. The observed fluxes present a modulation depending on the solar longitude. The bottom panel presents the averaged flux measured by both spacecraft as a function of the solar ecliptic longitude, with the vertical line indicating the upstream direction of the interstellar dust flow, and the solid line showing the best fit to the data of a simple model for orbital effects.

### 3.2.4 EXPLORATION SCIENCE

**Lunar L2-Farside Missions with the Orion Multi-Purpose Crew Vehicle (MPCV)**
*Investigators:* Jack Burns, U. Colorado; Scott Norris & Josh Hopkins, Lockheed-Martin; D. Kring, LPI; J. Lazio, JPL

The LUNAR team, including our corporate partner Lockheed-Martin, is working on Exploration Science using the Orion MPCV. Exploring the Moon's farside is a possible early goal for missions beyond Low Earth Orbit using the Orion spacecraft to explore incrementally more distant destinations. The lunar L2 Lagrange Point is a location where the combined gravity of the Earth and Moon allows a spacecraft to be synchronized with the Moon in its orbit around the Earth, so that the spacecraft is relatively stationary over the farside of the Moon.

The farside has been mapped from orbit but no humans or robots have ever landed there. There are two important science objectives on the farside. The first is to return to Earth multiple rock samples from the Moon's South Pole–Aitken basin which is one of the largest, deepest, and oldest craters in the solar system. The second objective is to deploy a low-frequency radio telescope on the farside where it would be shielded from human-generated low frequency radio interference, allowing astronomers to explore the currently unobserved Dark Ages and Cosmic Dawn. These early Universe observations were identified as one of the top science objectives in *New Worlds, New Horizons in Astronomy and Astrophysics*: "Cosmic Dawn: Searching for the First Stars, Galaxies, and Black Holes."

> "The exploration and sample return from the Moon's South Pole-Aitken basin is among the highest priority activities for solar system science…A robotic lunar sample return mission has extensive feed forward to future sample return missions from other locations on the Moon as well as Mars and other bodies in the solar system."
> *Vision & Voyages for Planetary Science in the Decade 2013-2022*

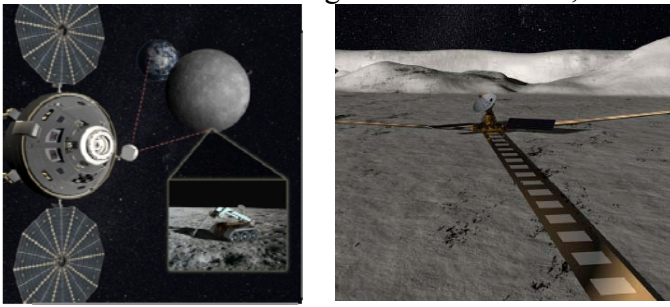

*Left:* Concept for the L2 mission teleoperating a rover on the farside of the Moon via the Orion Crew Module. *Right:* An artist impression of our low frequency array deployed on the farside of the Moon.

The L2-Farside mission is a logical early step beyond low Earth orbit in advance of longer trips to more distant and challenging destinations like asteroids and eventually Mars. The LUNAR team has collaborated with Lockheed-Martin in developing a mission scenario in which astronauts, orbiting at the unique vantage point of L2, could remotely operate robots on the lunar surface to deploy a low frequency farside array of radio antennas. Astronauts on an L2-Farside mission would travel 15% farther from Earth than the Apollo astronauts did, and spend almost three times longer in deep space. Each flight would prove out Orion's life support systems and would demonstrate the high-speed reentry capability needed for return from the Moon or deep space – 40 to 50% faster than reentry from low Earth Orbit.

## LUNAR Simulation Laboratory at the University of Colorado
*Investigators:* Jack Burns with students  Kristina Davis, Laura Kruger, Chris Yarrish, U. Colorado

The LUNAR team has constructed a LUNAR Simulation Laboratory (LSL) at U. Colorado to mimic lunar surface conditions during the day and night, and to test materials and material deployment that might be used for future space science and human exploration of the Moon. The core of the LSL consists of a vacuum chamber with a diameter of 1.1 ms and depth of 0.9 m, and a vacuum pressure of $10^{-8}$ torr.  The simulated lunar surface plate can be thermally cycled to temperatures 100 C to -150 C, equivalent to that on the Moon.  An ultraviolet lamp simulates the photon radiation environment of the Moon down to $\lambda \approx 100$ nm.    A video camera monitors experiments in real time. The thermal, mechanical, and electrical properties of test materials within the chamber can also be continuously measured.

LUNAR has carried out experiments to investigate the feasibility of using Kapton, a polyimide film, as the framework for a lunar telescope array. The telescope will be subjected to harsh conditions including 250 C variations in temperature and exposure to ultraviolet radiation. Our experiments have investigated any changes in the Kapton's properties. Most recently, two experiments were performed each lasting one month.

> "The lunar surface offers extraordinarily radio-quiet sites on the lunar farside that could enable a highly sensitive low-frequency radio telescope…An innovative concept recently proposed would have a complete antenna line electrodeposited on a long strip of polyimide film…As a result of the high astronomical priority of this work and the uniquely enabling character of the radio-quiet farside lunar surface, such efforts deserve cultivation."
> *The Scientific Context for the Exploration of the Moon*

The vacuum chamber used was taken down to high vacuum ($<10^{-7}$ Torr) and cycled down to -150 C for 24 hours and then up to 100 C for 24 hours. This was repeated for one month, simulating one lunar year. After analyzing data from this experiment, it appears that Kapton is a viable backbone material for a radio array.

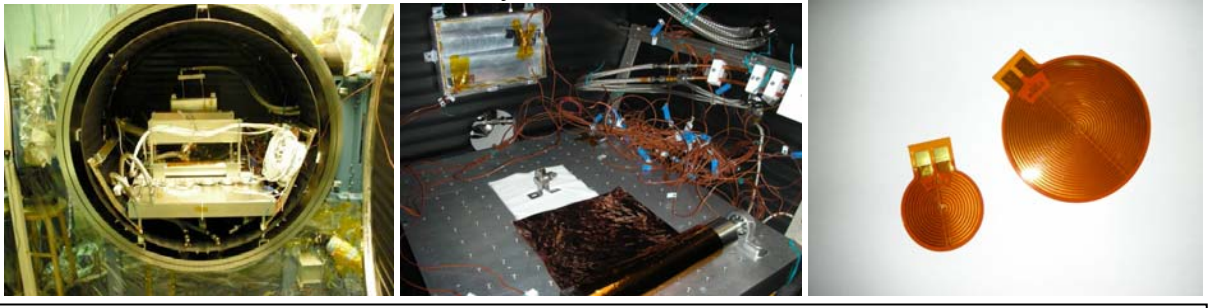

*Left:* The Lunar Simulation Laboratory vacuum chamber. *Center:* a roll of copper-coated Kapton film has been mechanically unrolled, exposed to repeated thermal cycling from -150 C to 100 C, and exposed to hard ultraviolet radiation during the "day" cycle. *Right:* Kapton films with embedded electrical circuits are shown which mimic the simplest components of a radio telescope array.

In the second test, a Kapton heater was used to simulate the electronics associated with dipole antennas. Analyses of the results of this experiment do not indicate any degradation of the electrical properties of the embedded circuit. The LUNAR team has also begun to work with Chris McQuin from the ATHLETE Rover Mechanical Team at JPL to do thermal testing of actuator/joint components of this candidate lunar rover using the LSL.

The LSL is designed to mimic the temperature and photon radiation environment of the Moon's surface over the course of a full lunar rotation. It is an appropriate facility to test science equipment and deployment techniques that are envisioned for both robotic and human exploration of the Moon as well as other bodies such as NEOs. Thus, the LSL is relevant to advancing the Exploration goals of NASA.

### Deployment Mechanisms for Surface Experimental Packages
*Investigators:* Robert MacDowall, GSFC and Joseph Lazio, JPL

During *Apollo*, astronauts deployed small instruments called *Apollo Lunar Surface Experiments*

*Package* (ALSEP). The return of robotic and crewed missions will be significantly enhanced if they are also capable of deploying small instrument packages, sometimes at distances from the lander; indeed, it is possible that automatic deployment of packages may be required as part of robotic precursor missions. One obvious deployment mechanism would be via a rover. LUNAR is developing an automatic deployment mechanism that is less massive and lower power. The deployment of a novel antenna concept for use in studying the surface boundary exosphere of an airless body such as the Moon or a precursor to a future solar radio observatory is being considered. Radio astronomy observations from the lunar surface provide access to data not available from Earth or satellites because of the lack of a significant lunar ionosphere, the stable 2-D surface on which to deploy antennas, and/or the radio-quiet zone on the lunar farside. Imaging of solar radio bursts at frequencies below those observable from Earth's surface would be possible for the first time with an array of ~50 antennas with a baseline of up to 1 km located on the lunar *nearside*. This concept has been proposed by the LUNAR team as the Radio Observatory on the Lunar Surface for Solar studies (ROLSS), where the antennas and leads are metal deposited on *rolls* of Kapton film. One or a few antennas on the lunar surface could also be used to track the density of the lunar ionosphere by using the technique of *riometry*. By measuring the cutoff frequency of a lunar ionosphere, a riometer uniquely determines the density of the ionized atmosphere of the Moon.

This deployer system, shown to the right, uses a spring deployed anchor to pull out and hold a double line, one end of which is attached to the leading edge of the film roll, the other to a small motor. In this model, the anchor and film canister are mounted on the outside of the lander. For deployment, the cover is opened, then the spring-driven anchor is released. Next, the motor begins pulling in the line, which sets the anchor and pulls out the film with the deposited antennas. This system will be capable of deploying other "packages," representing small devices like electrometers or biosensors, thereby demonstrating a range of applications for the deployer.

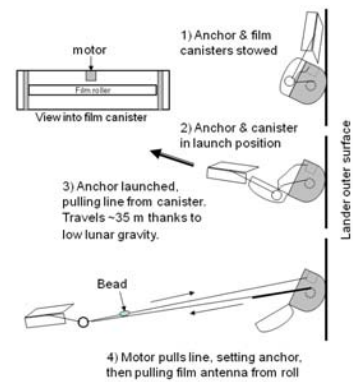

As a risk reduction step for human exploration, it may be necessary to undertake robotic precursor missions. If so, it is possible that automatic deployment of packages such as proposed by the LUNAR team may be required as part of these missions.

Also being considered are various options for lunar antenna deployment via inflatable tubes, which would be allowed to deflate after deployment. The current assessment is that this method is likely to be a viable approach for arm lengths up to ~ 50 meters; not yet clear is how much longer this technique could be made to work reliably.

### New Approaches to Drilling in the Lunar Regolith for Lunar Laser Ranging
*Investigators:* Douglas Currie, U. Maryland and Kris Zacny, Honeybee Robotics

Many of the planned lunar science activities require drilling into the regolith, for example heat flow, core sampling as well as the optimal emplacement of next generation Lunar Laser Ranging Retroreflectors (LLR). New technologies have emerged since the *Apollo* missions but have not been tested "in-situ". LUNAR team members are working with *Astrobotics* and *Honeybee* in order to define a new lunar laser ranging system as part of the Google Lunar X Prize (GLXP) program. This will consist of using the pneumatic drill developed by *Honeybee* on the Lander being developed by *Astrobotics*. Thus, this whole new concept may be tested in the real environment within the next few years. This program will provide information that will be of great value for missions like the Lunar Geophysical Network (a.k.a. ILN) as well as other missions to land on the surface of airless bodies.

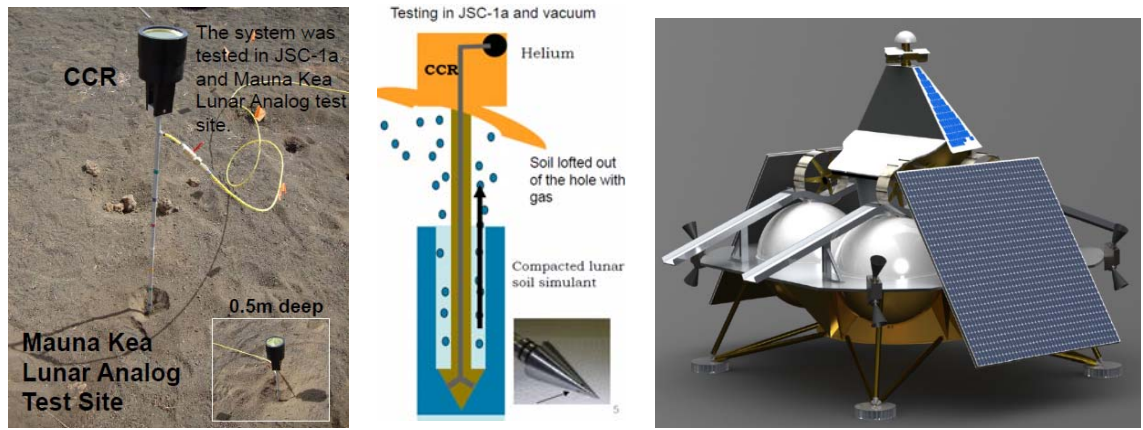

*Left &* Center: Gas assisted drilling tool developed by Honeybee Robotics with support from LUNAR. *Right:* ***Astrobotics*** lander, including the Honeybee drill, is being proposed to deploy a next generation lunar laser retroreflector as part of the Google Lunar X-Prize competition.

The drilling technology developed to deploy a next generation LLR will be fundamental to construction and science activities as part of human exploration of the Moon. The *Apollo* astronauts demonstrated the difficulty of drilling and securing instruments to the lunar regolith using traditional tools. The drilling technology development supported by LUNAR will revolutionize human surface operations in the lunar environment.

### Investigation of the Mechanical Properties of Lunar Regolith Simulant Cements
*Investigators:* Peter Chen and Douglas Rabin, GSFC

The surface of the Moon is covered by a layer of dust several meters deep. This lunar regolith is generally considered a nuisance and a health hazard. However, research undertaken LUNAR is studying a variety of ways to put this regolith material to use for the enhancement and enablement of future activities on the Moon and other airless, dust-coated bodies.

We have found that, using a mixture a lunar regolith simulant (JSC-1AC), an epoxy, and carbon nanotubes, a very hard substance akin to cement is formed. This 'lunar cement' can be used e.g., to fabricate solar collectors as well as large telescopes. LUNAR is investigating new types of regolith simulants, the mechanical properties of the cement, and the development of applications that are unique to the Moon.

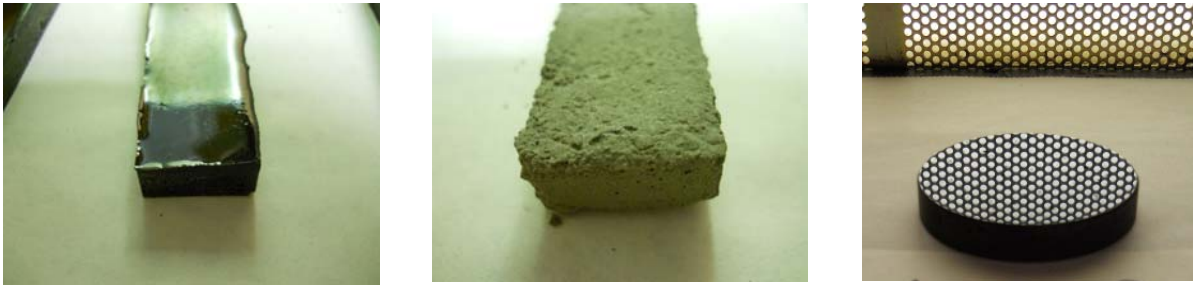

*Left & Center:* Samples of 'lunar cement' made with lunar regolith simulant and epoxy at different proportions. The ratio of simulant to epoxy that could make a robust material turned out to be quite narrow. On the left is a 3:1 sample. At 5:1 (center), the material was crumbly and had very low structural strength. *Right:* A 5 cm diameter lunar cement disk was polished into an optical flat. The experiment suggests that future telescopes on the Moon can be made in-situ using lunar dust.

An important goal for human exploration of the solar system involves learning to "live off the land" on planetary surfaces. The work of our LUNAR team to fabricate hardened structures from regolith simulant cement has potential wide applications beyond future telescopes, including solar arrays, habitats, roadways, and radiation shielding.

### Selenodesy

*Investigators:* Thomas Murphy, UCSD; D. Currie, U. Maryland; S. Merkowitz, GSFC

Lunar Laser Ranging (LLR) is one of the key projects for the LUNAR team. LLR provides unique measurements of the Moon's fluid core as well as constraints on gravitation via limits on the deviations from General Relativity. Recently, a major advance in LLR was made via the recovery of the Soviet-era Lunokhod 1 retroreflector. Lunokhod 1 (figure below) has never been ranged to, since the coordinates of the final resting place were not sufficiently accurate. Using high resolution imagery, LRO (LROC) identified the location of Lunokhod 1 and these coordinates were given to LUNAR Co-I Murphy for use at the APOLLO (Apache Point Observatory Lunar Laser ranging Operation) Station. This allowed laser returns to be obtained from Lunokhod 1 for the first time. However, the LLR position of Lunokod 1 was different from the LRO coordinates by 100 meters. Thus, this new tie point should allow a very significant upgrade to the selenodetic coordinate system being used by LRO. Additional retroreflectors, which the LUNAR team is designing, will serve to tie down the coordinate system in a variety of new locations.

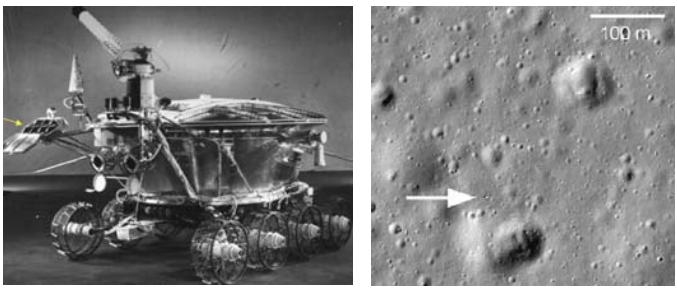

An accurate grid or selenographic coordinate system will be critical in future robotic and/or manned missions. The LUNAR LLR program, using Apollo-era retroreflectors along with next-generation retroreflectors emplaced robotically, will provide a lunar coordinate system accurate to 1-meter that will be essential for human activities on the surface of the Moon.

*Left:* Lunokhod 1 lander. *Right:* LRO LROC image of the Lunokhod 1 lander.

## 3.3　Inter-Team Collaborations and International Partnerships

From the inception of LUNAR, we have focused on coupling our science to exploration. Bridging these two will be pivotal in the development and deployment of various scientific instruments and experiments. We have strong scientific partnerships with NLSI teams CCLDAS (Mihaly Horanyi), DREAM (Bill Farrell) and CLSE (David Kring). We have also reached out internationally and are working with groups in Europe and Japan.

### Lunar Laser Ranging (LLR)

A project was initiated in order to evaluate the effect of micrometeorite impacts as an explanation of the reduced return from the Apollo arrays. This is critical in order to implement design aspects for the LLRRA-21 to assure a long lifetime with high return signal strength. To this end, witness sample plates were bombarded with accelerated dust particles in the CCLDAS dust accelerator at U. Colorado. This resulted in some cratering by dust particles of the highest energy, but given the energy vs. number information, this does not seem to be the dominant cause. This now appears to be dust, either secondary ejecta or lofted dust.

Discussions with the DREAM team are continuing to address the expected trajectories of lofted dust. We are reviewing these dust models to determine how dust might affect the LLRRA-21 (i.e., next generation Lunar Laser Retroreflector) design.

A productive collaboration has been developed between LUNAR members at U. Maryland and the INFN-LNF in Frascati, Italy. LUNAR is using the thermal vacuum facility at INFN-LNF to test designs for the optical and thermal properties of LLR corner cube reflectors and retroreflector arrays. The resulting thermal analyses and simulations of performance are strongly influencing the candidate LLRRA-21 designs.

The National Astronomical Observatory of Japan (NAOJ) has been working with our LUNAR LLR analysis program to evaluate the parameters of the science to be achieved with the retroreflector that is proposed for SELENE-2. An agreement between LUNAR via U. Maryland, the NAOJ, JAXA, the INFN-LNF, and the Institut d'Astrophysique Spatiale (IAS) has been signed to extend this to an evaluation of expected science accuracy and time scales for a 1-mm ranging accuracy from multiple ground stations.

### Low Frequency Cosmology and Astrophysics (LFCA)

The LUNAR Low Frequency Cosmology & Astrophysics project team has been interacting with other NLSI teams, most notably the Dynamic Response of the Environment At the Moon (DREAM) team. The most significant collaborations have involved the Lunar Radio Array; specifically, we have worked with Bill Farrell on the impact of the lunar environment (solar wind) on future lunar radio telescope antennas on the lunar surface. Farrell (DREAM) *et al.* presented a poster, on behalf of the DREAM and LUNAR teams, at the 2010 Lunar Science Forum reporting on a preliminary investigation of the plasma environment on the performance of polyimide film-based lunar radio antennas ("Lunar Environmental Effects and Astrophysical Platforms").

Lazio and Farrell (DREAM) published a paper on radio observations of the extrasolar planet HD 80606b.  While they were unable to detect it, they showed that the most likely frequency at which this planet would emit would be below 100 MHz, which is in the likely frequency range for a future lunar cosmology radio telescope (Lazio et al. 2010).

The LFCA project team has collaborated with researchers in Europe on international concepts for low frequency telescopes in space and on the Moon.  In particular, we have worked with LUNAR international collaborator Dr. Heino Falcke (Max Planck Institute for Radioastronomy, Bonn) on concepts for deployment of low frequency antennas on the Moon using proposed designs of ESA lunar landers/rovers.

## Radio Heliophysics
Much of the pioneering work utilizing low frequency radio antennas to detect the impact of microscopic grains on spacecraft was done at the Paris Observatory. With the launch of the STEREO spacecraft in 2006, significant progress is underway in the explanation of dust properties using radio measurements due to the high sensitivity of the STEREO. LUNAR partnered with the Paris observatory to quantify how accurately a radio instrument on the Moon could survey interplanetary dust as a function of grain size and time. Paris Observatory researcher A. Zaslavsly was hired as a postdoc at SAO, where he developed algorithms for equating the waveform produced by a dust impact to the properties of the dust grain.  Zaslavsky has returned to Paris to accept a faculty position.

## Exploration Science
The LUNAR team, including our corporate partner Lockheed-Martin, is working on Exploration Science using the Orion MPCV. Exploring the Moon's farside is a possible early goal for missions beyond Low Earth Orbit using the Orion spacecraft to explore incrementally more distant destinations. The lunar L2 Lagrange Point is a location where the combined gravity of the Earth and Moon allows a spacecraft to be synchronized with the Moon in its orbit around the Earth, so that the spacecraft is relatively stationary over the farside of the Moon.

The farside has been mapped from orbit but no humans or robots have ever landed there. There are two important science objectives on the farside. The first is to return to Earth multiple rock samples from the Moon's South Pole–Aitken (SPA) basin which is one of the largest, deepest, and oldest craters in the solar system.  We are collaborating with David Kring's CLSE team to define the geological science objectives for a teleoperated rover controlled by astronauts aboard Orion.  The second objective is to deploy a low-frequency radio telescope on the farside where it would be shielded from human-generated low frequency radio interference, allowing astronomers to explore the currently unobserved Dark Ages and Cosmic Dawn.

## Education and Public Outreach
Locally in Boulder, the CCLDAS and LUNAR teams partnered to run a Galileoscope teacher workshop led by Erin Wood (CCLDAS) and Matt Benjamin (LUNAR).  Every teacher who attended received an inexpensive "Galileoscope" they could keep and practiced observing the moon.  Activities related to understanding cosmology and the role of observations from the moon were also taught.  The impressive "Science on a Sphere" at Fiske Planetarium was used to display lunar data.  A "NASA Nugget" was submitted about this meeting.

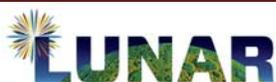

In association with the LCROSS/LRO team we held a "Lunar Bagel Breakfast" to watch the LCROSS impact event. More than 200 people filled the Fiske Planetarium Theater for this pre-dawn event, where we broadcast NASA TV and viewed CCD images from our 24-inch campus observatory. LUNAR scientists such as LUNAR PI Jack Burns provided commentary. We arranged for good local newspaper and TV coverage, and LUNAR Co-I Duncan also appeared on the NBC national news.

## 3.4    Education and Public Outreach (EPO)

*Leaders*: Douglas Duncan U. of Colorado, Matt Benjamin U. of Colorado

### Highlights

The Lunar University Network for Astrophysics Research (LUNAR) team has a diverse and aggressive Education and Public Outreach effort aimed at enhancing the awareness and knowledge about the Earth-Moon system. The largest elements of this effort are the creation of a nationally-distributed children's planetarium show and extensive teacher workshops, many in partnership of the Astronomical Society of the Pacific. A smaller element, but a highlight, is support for high school robotics clubs making their own models of a lunar rover capable of deploying a radio telescope antenna on the lunar surface. A final strategy is to take advantage of NASA missions and natural events such as eclipses to increase public awareness of science and of NASA's role.

The children's planetarium program is based on the award winning book, "Max Goes to the Moon" by local Boulder author Dr. Jeffrey Bennett. NASA astronaut Alvin Drew played a role in the development of this show. On his mission to the International Space Station he had the opportunity to read the story "Max Goes to the Moon" to the children of Earth. Because of that we asked him to introduce the story in the planetarium show. Using our well-developed process of "formative evaluation" (extensive beta-testing during development), we showed the program to test audiences of school children of the target age and also to hundreds of lunar scientists at the 2011 NLSI workshop. The feedback we gathered resulted in significant improvements to the show. "Max" is now complete and we are beginning distribution of the program.

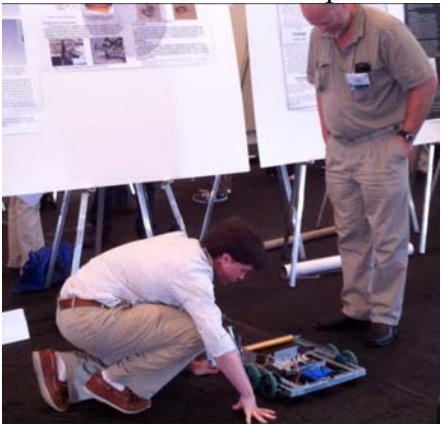

LUNAR high school student at the Lunar Science Forum 2011 about to demonstrate his rover to deploy a roll of polyimide film.

Our numerous K-12 teacher workshops focused on getting the latest discoveries about the Earth-Moon system and cosmology and the early Universe into the classroom. By holding workshops at the national Astronomical Society of the Pacific meeting, we increased the number of teachers who we reached to much higher numbers than we originally proposed.

LUNAR Co-I Lazio began working with high school robotics clubs to challenge them to build a small rover that could deploy low frequency radio telescopes on the near and far sides of the Moon. A student from one of these schools was invited to the NLSI Lunar Forum in the summer of 2011 to demonstrate his rover to the lunar community.

We planned public events associated with the LCROSS/LRO mission as well as the lunar eclipse of Dec. 2010. The lunar eclipse was the largest public astronomy event in Boulder, Colorado since the "Deep Impact" comet mission in 2005. Approximately 1500 people crowded into a planetarium that seats 212 (using the lobby, the grounds, and the surrounding university). All heard about NASA lunar science in addition to seeing the eclipse.

The LUNAR team website (lunar.colorado.edu) focuses not just on the science we are doing but the human element of space science and exploration, to be more involving to the reader. We created and developed a LUNAR team brochure, similar to that of the NLSI. Both have been successful ways to publicize space and lunar science, and the coupling of science and exploration.

### *"Max Goes to the Moon" Planetarium Show*
*Leaders*: Douglas Duncan, U. of Colorado, Jeffrey Bennett, U. of Colorado, Matt Benjamin, U. of Colorado

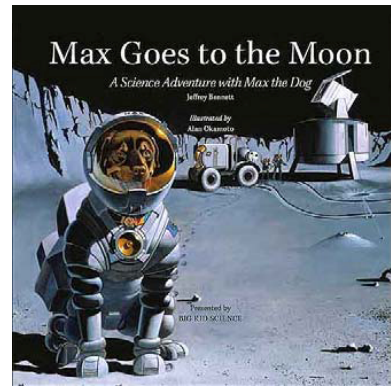

This is a children's show aimed at grades 2-5, based on the award-winning children's book by Dr. Jeffrey Bennett. Dr. Bennett is also one of the authors of the most widely used astronomy textbooks in the country, *The Cosmic Perspective*. We were able to take the book and animate it in a way that stayed true to the book while making the show more visually interesting. We also inserted 3D models and animations provided by Lockheed Martin Corporation and NASA, as well as making our own. Working with NASA astronaut Alvin Drew gave us the opportunity to record him in front of a green screen. This show is now completed and we have begun preparing this show for distribution across the US and World. We already have requests from 10 US planetariums and 3 international ones.

Originally we sought to make "Max" appropriate for all elementary grades, but our formative evaluation revealed that this was simply too wide a range in student development, and we sharpened the intellectual level to be appropriate to 2-5 graders, with extra vignettes designed to keep the interest of parents and teachers.

### Teacher Workshops

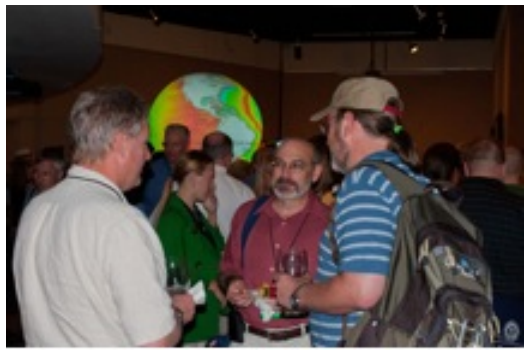

Our EPO plan called for running one teacher workshop during summer 2010, with attendance of 15 teachers. However, by joining with the Astronomical Society of the Pacific we became part of the national meeting, "Cosmos in the Classroom," which was held at the University of Colorado, with LUNAR EPO lead Doug Duncan as Chair of the Local Organizing Committee. This allowed us to draw 50 teachers and break them into elementary, middle school, and high school groups. The LUNAR team's Matt Benjamin and Doug Duncan taught at the workshops. As specified in our work plan, we created some new materials and also took advantage of good materials created by NASA and by other educators who participated in the ASP meeting.

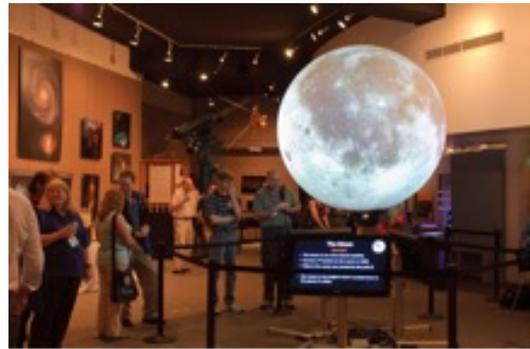

We also participated in the ASP's teacher workshops in 2009. A particularly successful activity was developed to explain the idea of cosmic "look-back time," the idea that the farther astronomers look into the universe, the older the light they are seeing, but the *younger* the universe looks. This is very important to LUNAR since we hope to see the universe before any stars or galaxies formed. The activity paired photos of older and older galaxies with old photos of the astronomer leading the activity. Students recognized that "old photos show young people" and that analogy was used to explain "old light shows young objects in the universe."

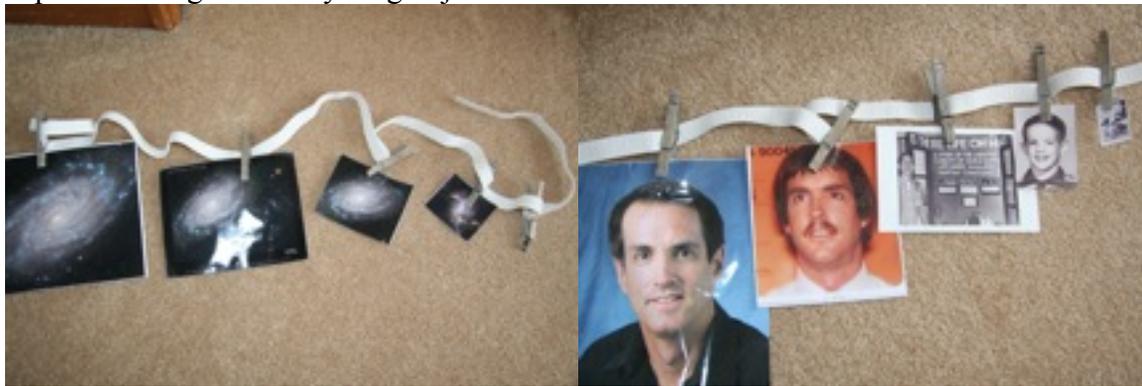

Locally in Boulder, the CCLDAS and LUNAR teams partnered to run a Galileoscope teacher workshop led by Erin Wood (CCLDAS) and Matt Benjamin (LUNAR). Every teacher who attended received an inexpensive "Galileoscope" they could keep and practiced observing the moon. Activities related to understanding cosmology and the role of observations from the moon were also taught. The very impressive "Science on a Sphere" at Fiske Planetarium was used to display lunar data. A "NASA Nugget" was submitted about this meeting.

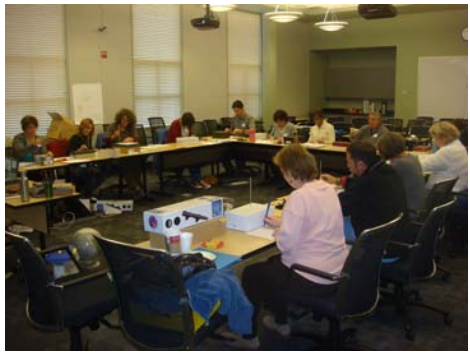
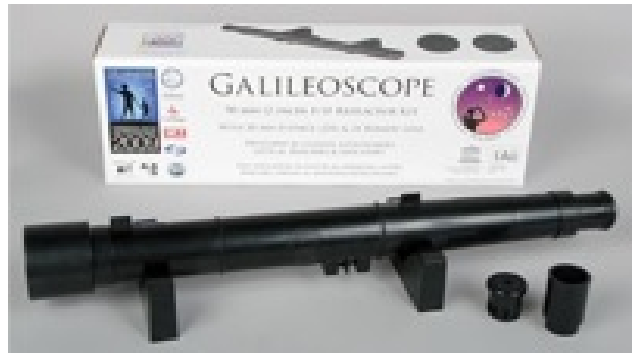

### Robotics Clubs Lunar Rovers

LUNAR Deputy Director Joseph Lazio initiated a project to work with high school robotics clubs to develop small remote-operated rovers. The goal is to get high school students to become excited by the opportunities available to the next generation of lunar scientists and engineers. Thanks to NLSI central, we received enough funds to support 3 schools. As of early 2011 we have request from 3 teachers and are buying Kapton film to distribute to them along with funds for parts to build robots. A student from one of these schools was invited to present a poster and his team's rover at the NLSI Lunar Science Forum in the summer of 2011. A video of the model robot achieved a considerable audience on YouTube.

## Community Outreach Events

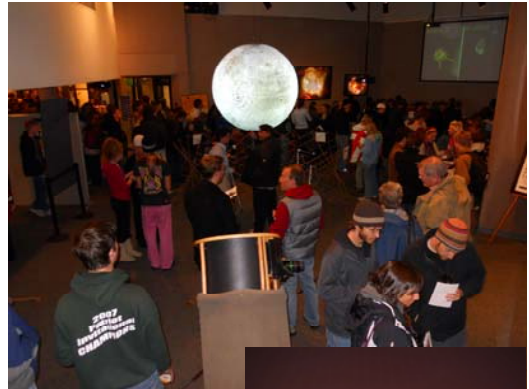

In association with The LCROSS/LRO team we held a "Lunar Bagel Breakfast" with food sponsored by local restaurant to view the LCROSS impact. More than 200 people filled the Fiske Planetarium Theater at this pre-dawn event, where we broadcast NASA TV, and others were in the lobby. LUNAR scientists such as Dr. Jack Burns provided commentary. We arranged for good local newspaper and TV coverage, and Fiske Director Duncan also appeared on the NBC national news.

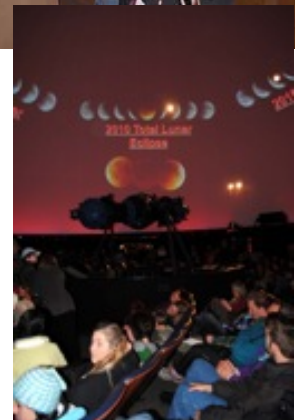

For the December 2010 lunar eclipse many telescopes were made available at Fiske Planetarium and Sommers Bausch Observatory at the University of Colorado, and local amateur astronomers were invited to set up more. TV stations were told about the eclipse well in advance, and all decided to cover it from Fiske, rather than Denver. The resulting crowd was roughly 1500 people. With clouds that cleared at just the critical time, everyone saw the eclipse. LUNAR team member Matt Benjamin presented a program in the planetarium theater that not only explained the eclipse but also highlighted NASA lunar science. He repeated the program three times in order to accommodate the large crowds.

## LUNAR Brochure

[Two-page brochure images]

The LUNAR team designed and printed its own brochure for use at workshops, conferences and EPO activities. We have been distributing these nationally and internationally.

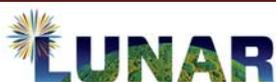

## 3.5 Peer-Reviewed Publications

Total Refereed Publications = *62*

Zeiger, B., & Darling, J. 2010, "Formaldehyde Anti-Inversion at $z = 0.68$ in the Gravitational Lens B0218+357," ApJ, 709, 386

Zacny, K., Currie, D., Paulsen, G., Szwarc, T., Chu, P., 2012, "Development and Testing of the Pneumatic Lunar Drill for the Emplacement of the Corner Cube Reflector on the Moon", submitted to *Planetary and Space Science*.

## *3.6 Conference Papers, Extended Abstracts, Posters, and Presentations*
Total of Conference Papers, Extended Abstracts, Posters, & Presentations = **220**

*Please go to this link to view all LUNAR abstracts submitted at first three years of the NLSI Lunar Science Forums:* ***http://lunar.colorado.edu/meetings/index.php***

### 3.6.1 Conference Proceedings, Reports, and Abstracts
Bale, S., 2010, "Solar and Interplanetary Radio Emission near the Moon", Robotic Science From the Moon Workshop. Boulder, CO

Benjamin, M., Burns, J., Duncan, D., 2009, "LUNAR EPO Planetarium Shows and Teacher Workshops", NLSI Lunar Science Forum.

Benjamin, M., Burns, J., Currie, D., Duncan D., Kasper, J., MacDowall, R., Lazio, J., 2010, "Year 1 Accomplishments", NLSI Lunar Science Forum.

Benjamin, M., Burns, J., Currie, D., Duncan D., Kasper, J., MacDowall, R., Lazio, J., 2011, "Year 2 Highlights for the NLSI LUNAR Team", NLSI Lunar Science Forum.

Boni, A., Dell'Angello, S., Currie, D., Alley, C., Arnold, D., Berardi, S., Bianco, G., Cantone, C., Delle Monache, G., Garattini, M., Intaglietta, N., Lops, C., Maiello, M., Martini, M., McGarry, J., Porcelle, L., Patrizi, G., Pearlman, M., Vittori, R., Zerbini, S., Zagwodzki, T., "World first SCF-Test of the NASA-GSFC LAGEOS Sector and Hollow Retroreflector" , (2011) International Laser Ranging Workshop, Bad Koetzting (GERMANY) 16 May – 20 May 2011

Bowman, J. 2009, "21 cm global signal: Earth-based constraints and implications for lunar observations," NLSI Lunar Science Forum.

Bowman, J. D., & Rogers, A. E. E. 2011, "Results From EDGES," Amer. Astron. Soc. 217[th] Meeting, Seattle, WA

Bowman, J. D., & Rogers, A. E. E. 2010, "VHF-band RFI in Geographically Remote Areas," Proceedings of Science

Bowman, J., 2010, "Lessons from EDGES", Robotic Science From the Moon Workshop. Boulder, CO

Taylor et al. 2011, "The Effect of Element Failures on an Optimized DALI Station", LUNAR memo No. 4

Traub, W. A., Lawson, P. R., Unwin, S. C., et al. 2010, "Exoplanets Forum 2008," in Pathways Towards Habitable Planets, eds. V. Coudé du Foresto, D. M. Gelino, & I. Ribas (San Francisco: Astron. Soc. Pacific) p. 21

Weiler, K. W., Lazio, J., Kasper, J., Burns, J., Jones, D. L., Furlanetto, S., MacDowall, R. J., Demaio, L., Bale, S., Ellingson, S., and Taylor, G. 2010, "The Dark Ages Lunar Interferometer (DALI)," in Astrophysics 2020: Large Space Missions Beyond the Next Decade, ed. W. Jin, I. Platais, and M. Perryman (Cambridge Univ. Press), presentation available at http://www.stsci.edu/institute/itsd/information/streaming/archive/Astrophysics2020

Yan, T. Stocke, J., & Darling, J. 2010, "Detection of Two Redshifted H I Absorption Systems," Amer. Astron. Soc. Meeting 215, #460.07

Zeiger, B. R., & Darling, J. 2010, "$H_2CO$ Absorption of the Cosmic Microwave Background: A Distance-Independent Tracer of Dense Molecular Gas," Amer. Astron. Soc. Meeting 215, #415.04

Zacny, K.; Currie, D.; Paulsen, G.; Avanesyan, A.; Chu, P.; Makai, T.; Szwarc, T. "Development and Testing of Gas Assisted Drill for the Emplacement of the Corner Cube Reflector System on the Moon", Annual Meeting of the Lunar Exploration Analysis Group, LPI Contribution No. 1646, p.85 11/2011

### 3.6.2 Colloquia, Public Presentations, and Posters

Bowman, J. "MWA Status and New Science," The Path to SKA-low Workshop, Perth, Australia, Sept. 6-9, 2011 (invited, session chair)

Bowman, J. "Overview of 21m Observables," Conference on Novel Telescopes for 21 cm Cosmology, Penticton, Canada, June 14, 2011 (invited review)

Bowman, J. "EDGES," Conference on Foregrounds for CMB and 21 cm, Zadar, Croatia, May 23-27, 2011 (invited, session chair)

Bowman, J. "Experiment to Detect the Global EoR Signature: Overview and Early Science," Hydrogen Cosmology Workshop, Harvard, May 18-19, 2011 (invited, session chair)

Bowman, J. "Results from EDGES," Abstract #107.08, Special Session on Hydrogen Epoch of Reionization Arrays (HERA), 217th Annual Meeting of the American Astronomical Society, January 9-14, 2011, Seattle, WA (invited)

Bowman, J. "Hydrogen Epoch of Reionization Array (HERA)," URSI-USNC, Special Session on Large-N Radio Arrays: Issues and Algorithms, Boulder, CO, January 5-8, 2011 (invited)

Burns, J. O.  2009 July, "Exploring the Cosmos from the Moon", NLSI Lunar Science Forum 2009 plenary science talk, NASA/Ames Research Center

Burns, J.  2010, "Exploring the Cosmos from the Moon," scientific colloquium, Naval Research Laboratory, Washington, DC

Burns, J.  2010, "Exploring the Cosmos from the Moon," scientific colloquium, MichiganState University, East Lansing, MI

Burns, J. 2010, "Exploring the Cosmos from the Moon," scientific colloquium, University of Michigan, Ann Arbor, MI

Burns, J.  2010, "Exploring the Cosmos from the Moon," invited public lecture, Aspen Winter Conference on The High Redshift Universe, Aspen, CO

Burns, J.  2010, "Exploring the Cosmos from the Moon,"  invited public lecture, Florida Institute of Human & Machine Cognition, Pensacola, FL

Burns, J.  2010, "Exploring the Cosmos from the Moon," invited public lecture, Conference on Nonthermal Phenomena in Colliding GalaxyClusters, Nice, France

Burns, J. 2011, "Exploring the Cosmos from the Moon", invited public lecture at the Florida Institute of Human & Machine Cognition, Ocala, FL, 23 February 2011.

Burns, J. 2011, "The Dark Ages Radio Explorer", presented as a CASA Astrophysics Seminar, University of Colorado, 18 January 2011

Burns, J. 2011, "The Dark Ages Radio Explorer", colloquium presented at the University of Waterloo, Canada, 24 March 2011

Burns, J. 2011, "Exploring the Cosmos from the Moon", invited public lecture at the University of Waterloo, Waterloo, Canada, 24 March 2011.

Burns, J. 2011, "The Dark Ages Radio Explorer", colloquium presented at the NASA Jet Propulsion Laboratory, Pasadena, CA, 11 April 2011

Burns, J. 2011, "The Dark Ages Radio Explorer", colloquium presented at the University of New Mexico, Albuquerque, NM, 28 April 2011

Burns, J. 2011, "The Dark Ages Radio Explorer", colloquium presented at the National Radio Astronomy Observatory, Socorro, NM, 29 April 2011

Burns, J. 2011, "The Dark Ages Radio Explorer", colloquium presented at the Max Planck Institute for Radio Astronomy, Bonn, Germany, 5 October 2011

Burns, J. 2012, "Highlights of LUNAR in Year 3", LUNAR webinar and seminar presented at the University of Colorado, 20 January 2012

Burns, J. O. 2012, "The Dark Ages Radio Explorer (DARE)," International Union of Radio Science-U.S. National Committee (Boulder, CO)

Chynoweth, K. M., Lazio, J., & Helmboldt, J. 2012, "A Constraint on the 21-cm Signal at z=20 from VLA Observations," International Union of Radio Science-U.S. National Committee (Boulder, CO)

Czekala, I., & Bradley, R. 2010, "Calibrating Astronomical Antenna Arrays with Man Made Satellites," American Institute of Aeronautics & Astronautics (AIAA), Region I-MA Student Conference, Blacksburg, VA

Czekala, I., & Bradley, R. 2010, "Calibrating Astronomical Antenna Arrays with Orbiting Data Satellites, Atlantic Coast Conference (ACC) "Meeting of the Minds" Conference, Atlanta, GA

Currie, D. G., & the LLRRA-21 Teams 2011 "A LUNAR LASER RANGING RETRO REFLECTOR ARRAY for the 21st CENTURY" 2nd Lunar Laser Ranging Workshop, International Space Sciences Institute, Bern Switzerland

Currie, D. G. "Lunar Laser Ranging Retroreflector Array for the 21st Century" 4th Science Conclave 2011- A Congregation of Nobel Laureates and Eminent Scientists, Indian Institute of Information Technology, Allahabad, India 11/2011

Currie, D., Dell'Agnello, S., Delle Monache, G., "Lunar Laser Ranging Retroreflector for the 21st Century", 17th International Workshop on Laser Ranging, Proceedings of the conference held 16-20 May, 2011 in Bad kotzing, Germany. To be published online at http://cddis.gsfc.nasa.gov/lw17

Currie, D., "Ground Stations for the Next Generation Lunar Retroreflectors", 17th International Workshop on Laser Ranging, Proceedings of the conference held 16-20 May, 2011 in Bad Kotzing, Germany. To be published online at http://cddis.gsfc.nasa.gov/lw17

Darling, J. "Hydrogen 21 cm Absorption Line Searches and Studies with SKAMP," Science with SKAMP: Widefield Spectroscopy of the Southern Sky, Sydney, Australia

Darling, J. "Redshifted OH Lines with SKAMP: Detection and Science," Science with SKAMP: Widefield Spectroscopy of the Southern Sky, Sydney, Australia

Darling, J. 2011 October, "Mining the Sky with WISE: Extreme Starbursts Spoofing H I and Other Oddities," Contributed talk for "Through the Infrared Looking Glass: A Dusty View of Galaxy and AGN Evolution" meeting, Pasadena, CA

## 3.7 List of Undergraduate Students, Graduate Students, Postdoctoral Fellows, and New Faculty involved in the LUNAR Team

*Numbers next to students indicate publications and presentations from Section 3.6.*

### University of Colorado Boulder
- Undergraduate Students
  - Riccardo Alfaro-Contreras    1
  - Kristina Davis    2
  - Katherine Grasha
  - Laura Kruger    3
  - Erin  Macdonald    1
  - Christopher Yarrish    1
- Graduate Students
  - Adrienne Dove (crossed-trained with Mihaly Horanyi's team)
  - Harrison Fast    2
  - Francesca Lettang    1
  - Jordan Mirocha    5
  - David Schenck
  - Samuel Skillman    6
  - Kyle Willett
  - Ting Yan    1
  - Benjamin Zeiger    3
- Postdoctoral Fellows
  - Abhirup Datta (NASA NLSI Postdoctoral Fellow)
  - Geraint Harker
  - Stephen Skory (25% NLSI funding)

### Arizona State University
- Undergraduate students
  - Hamdi Mani
  - Sarah Easterbrook
  - Jose Chavez-Garcia
- Graduate students
  - Jacqueline Monkiewicz
  - Thomas Mozdzen

### Harvard University
- Graduate Students
  - Jonathan Bittner    2
  - Bennett Maruca

- o Eli Visbal    4
- o Peter Adshead    1
  - Postdoctoral Fellow
    - o Jonathan Pritchard (Hubble Fellow)

### Harvard-Smithsonian Center for Astrophysics
- Postdoctoral Fellows
  - o Gaetan Le Chat
  - o Arnaud Zaslavsky

### Massachusetts Institute of Technology
- Undergraduate Students
  - o Rurik Primiani

### University of California at Los Angeles (UCLA)
- Undergraduate Students during summer internship at UCLA
  - o Sarah Benjamin (Carnegie Mellon)
  - o Samuel Johnson Stoever (Cornell)    1
- Graduate Students
  - o Lauren Holzbauer    1
- Postdoctoral Fellows
  - o Joseph Munoz

### Princeton University
- Postdoctoral Fellow
  - o Andrei Mesinger

### University of New Mexico
- Graduate Students
  - o Cristina Rodriguez

### University of Virginia
- Undergraduate students
  - o Ian Czekala    2

### Thomas Jefferson High School for Science & Technology (Virginia)
- S. Carmichael    2
- J. Clark    2
- E. Elkins    2
- Peter Gudmundsen    2
- Zachery Mott    2
- Melanie Szwajkowski    2
- Nathaniel Shkolnik    1